\documentclass[a4paper,11pt]{article}
\usepackage{tgtermes}
\usepackage{authblk}
\usepackage{amsmath}
\usepackage[utf8]{inputenc}
\usepackage{pdfpages}
\usepackage{graphicx}
\usepackage{physics}
\usepackage{amsmath}
\usepackage{epstopdf}
\usepackage{amssymb}
\usepackage{mathtools}
\usepackage{amsthm}

\usepackage[toc,page]{appendix}

\usepackage[colorlinks]{hyperref}
\AtBeginDocument{%
  \hypersetup{
    citecolor=blue,
    linkcolor=blue,
    urlcolor=blue}}

\usepackage[a4paper, total={6.4in, 9in}]{geometry}
\providecommand{\keywords}[1]
{
  \small	
  \textbf{Keywords---} #1
}

\theoremstyle{definition}
\newtheorem{defi}{Definition}[section]
\begin{document}

\title{Practically feasible robust quantum money \\ with  classical verification}

\author{Niraj Kumar \thanks{Email id- nkumar@exseed.ed.ac.uk}} \affil{School of Informatics, University of Edinburgh}
\date{\today}
\maketitle

\begin{abstract}
We introduce a private quantum money scheme with the note verification procedure based on Sampling Matching, a problem in one-way communication complexity model introduced by Kumar et al. \cite{kumar2019experimental}. Our scheme involves a Bank who produces and distributes quantum notes, noteholders who are untrusted, and trusted local verifiers of the Bank to whom the holders send their notes in order to carry out transactions.
The key aspects of our money scheme include: note verification procedure requiring a single round classical interaction between the local verifier and Bank; fixed verification circuit that uses only passive linear optical components; re-usability of each note in our scheme which grows linearly with the size of note; and an unconditional security against any adversary trying to forge the banknote while tolerating the noise of upto 21.4$\%$. We further describe a practical implementation technique of our money scheme using weak coherent states of light and the verification circuit involving a single 50/50 beam splitter and 2 single-photon threshold detectors. Previous best-known matching based money scheme proposal \cite{amiri2017quantum} involves a verification circuit where the number of optical components increase proportional to the increase in desired noise tolerance (robustness). In contrast, we achieve any desired noise tolerance (upto a maximal threshold value) with only a fixed number of optical components. This considerable reduction of components in our scheme enables us to reach the robustness values that is not feasible for any existing money scheme with the current technology.
\end{abstract}


\normalsize

\section{Introduction}
In the 1980s, Wiesner \cite{wiesner1983s} proposed the idea of quantum money to create unforgeable banknotes with quantum states. In his scheme, the banknotes are several BB84 states prepared by an honest authority, Bank, who then distributes them to the untrusted holders. When the holders need to carry out a transaction with their note, they send the entire note to the Bank for verification, who declares its validity. The unforgeability property of the note in Wiesner's scheme relies on the no-cloning property of quantum mechanics which prevents the holder from creating multiple copies of the notes with just a single copy \cite{wootters1982single}. This idea was incidentally also among the first quantum cryptographic primitives to be introduced. Subsequently, other cryptographic tasks based on quantum mechanics have been proposed such as quantum key distribution, digital signatures, coin flipping, secure multi-party computation, etc \cite{bennett2014quantum, gottesman2001quantum,ambainis2004new, broadbent2009universal, crepeau2002secure, broadbent2016quantum}.  

This Wiesner money scheme, as analysed independently by Lutomirski et al. \cite{lutomirski2010online} and Brodutch et al. \cite{brodutch2014adaptive}, soon ran into security problems. The first issue was $(a)$ verification of the note required  a quantum communication channel between the holder and the Bank. As pointed out by Gavinsky \cite{gavinsky2012quantum}, an adversary can interfere in the channel and possibly modify or destroy the note; $(b)$ the scheme is insecure against several previously un-analysed attacks, the adaptive attacks \cite{brodutch2014adaptive} where an adversary can substantially increase the note forging probability by communicating with the Bank in a few ``auxiliary" number of rounds.

These two drawbacks were first addressed by Gavinsky \cite{gavinsky2012quantum} in his proposal of a private quantum money scheme based on quantum retrieval games (QRG). Informally,  these games consist of a set of challenges where answering a single challenge from the set is easy, however answering multiple challenges simultaneously is hard i.e. cannot be answered with a unit probability. Gavinsky's scheme involves an honest Bank that prepares and distributes notes which are quantum states prepared some random basis, untrusted noteholders and trusted local verifiers of the Bank who run a test on holder's note to check for its validity. 
The security of this scheme relies on the fact that an honest user must always pass the verifier test who picks a single challenge from the set at random. A dishonest user who wants to forge the note has to simultaneously succeed in answering any two challenges picked from the set at random by two independent verifiers. Thus the hardness of forging the banknote links to the hardness in answering any two different challenges picked at random from the set. 
The verification of the note in Gavinsky's scheme requires two rounds of classical communication between the Bank and the verifier and is proven to be secure against any type of adaptive adversary attacks. This scheme, however, is not realistic since it only works when no experimental noise is taken into consideration. Also, since the verification procedure requires two rounds of communication, this forces the Bank to have a temporary ``active" classical memory during the verification phase which would limit the number of independent note verifications that the Bank can perform at any given instant. 

Further independent works on similar lines by Georgiou et al. \cite{georgiou2015new} and Amiri et al. \cite{amiri2017quantum} have reduced the classical communication required for note verification to a single round. The scheme of \cite{georgiou2015new} is based on 1-out-of-2 QRG i.e. the holder can deterministically answer any one of the two challenges, however, it is impossible for him to answer both the challenges simultaneously with a unit probability.  This scheme tolerates the noise of up to $12.5\%$. The other scheme of \cite{amiri2017quantum} is based on Hidden Matching quantum retrieval games, HM-QRG \cite{gavinsky2007exponential, arrazola2016practical}, and exhibits the noise tolerance of up to $23.3\%$. They further conjecture that maximal noise tolerance for money schemes based on Matching QRGs can reach up to $25\%$. Here the noise tolerance, a measure of the robustness of the scheme, is defined as the maximum theoretical probability that an honest verifier returns an incorrect outcome. Higher the noise tolerance, more robust is the money-scheme against errors incurred on the honest holder's note due to experimental imperfections. And as long as the errors on the honest holder's note is within the noise tolerance, the money scheme would demonstrate an information-theoretic security against a forger trying to forge the banknote. In parallel, several other theoretical proposals of quantum money schemes have been proposed, both as private and public money schemes \cite{pastawski2012unforgeable, aaronson2012quantum, farhi2012quantum, moulick2016quantum}. Besides, there has been a recent proposal on semi-quantum money \cite{radian2019semi} where the Bank only has classical resources and delegates a classical message to the user who prepares the quantum note. The security of this scheme is however computational and relies on the security provided by learning with errors (LWE), a problem believed to be post-quantum secure.

To date, there have been two proof-of-principle experimental demonstrations for private quantum money based on one round of classical verification with the Bank. The first by Bozzio et al.  \cite{bozzio2018experimental} is based on the theoretical scheme of \cite{georgiou2015new}. Their encoding of quantum money is in polarized weak coherent states and achieves the honest noteholder error rate slightly below $\beta = 4\%$ which is well under that maximum noise tolerance of $12.5\%$. The other demonstration is by Guan et al. \cite{guan2018experimental} which is based on the theoretical scheme of \cite{amiri2017quantum} has an encoding based on phase parity of corresponding pairs of weak coherent states. Their implementation achieves a measured honest holder error rate of $\beta = 3\%$, while the theoretical noise tolerance of the scheme is $16.6\%$, thus making it secure. Inspite of the maximal noise tolerance of $23.3\%$ proposed in the theoretical scheme, the experimental scheme was limited to $16.6\%$. The reason for this being that for the protocols based on matching schemes, the tolerance against the noise increases with the input size of the note. Thus the money scheme becomes more robust against experimental imperfections and forging by going to higher input sized banknotes. For the schemes based on Hidden Matching, the verification protocol involves a complex circuit with the number of optical elements (active switches, delays, beam splitter) increasing at least logarithmically with the input size. This gets increasingly difficult to implement the circuit for large input sizes. Hence the only implementation based on Hidden Matching has been shown for input size $n = 4$ which leads to a noise threshold of $16.6\%$. This is the primary motivation of our theoretical work which simplifies the verification circuit to be able to experimentally achieve much higher noise tolerance than any other current private money scheme.

In this work, we introduce a private quantum money scheme using single-photon quantum states and the verification protocol based on Sampling Matching (SM) scheme. This scheme was proposed by Kumar et al. \cite{kumar2019experimental} as a problem in a one-way randomized communication complexity model. The authors showed that solving this problem using quantum resources exhibits an exponential reduction in communication compared to classical resources. They further experimentally demonstrated the quantum protocol using weak coherent states and linear optics operations. The simplicity of the problem allowed for the construction of such a quantum protocol consisting of only $\mathcal{O}(1)$ 50/50 beam splitters and two threshold detectors independent of the problem size. This led to the first experimental realization of quantum advantage in one-way communication complexity. This problem is described in more detail in the section~\ref{tools}. 

We propose two distinct encoding of private quantum money schemes. The first proposal uses single-photon states to create a quantum note. The single-photon encoding is the realisation of a qubit/qudit with linear optics. Thus defining our money scheme in this picture makes it more translatable to other qubit encoding pictures for experimental demonstrations. Our money scheme, while achieving a noise tolerance of $21.4\%$, offers much simplicity in the implementation of the verification protocol compared to the existing Hidden Matching based protocols \cite{amiri2017quantum}. In our verification scheme with single-photon states, even though the number of linear optical components grows linearly with the input size, the optical components required are just passive 50/50 beam splitters with no need for active switch components. This allows our circuit to be fixed for a given input size of the note, in contrast to the scheme of \cite{amiri2017quantum} which requires a programmable verification circuit thus necessitating the need for having active components. 

In the second part, we describe a practical implementation technique of our money scheme using weak coherent states and the verification circuit consisting of a single 50/50 beam splitter and 2 single-photon threshold detectors. Here we show that it is experimentally possible to achieve a noise tolerance as high as $~21.4\%$, something not possible for any other scheme with the current technology.    

The paper is organized as follows. In section~\ref{defin} we define the private quantum money including the notions of correctness and unforgeability. In section~\ref{tools} we introduce the tools required to construct our money scheme. This includes defining a modified version of the SM problem  and SM scheme with single-photon states. In section~\ref{QuantumMoney} we formally introduce our quantum money scheme using SM verification. Sections~\ref{HonestHolder} and \ref{unfor} analyses the security of our money scheme and prove that it exhibits an information-theoretic security. Finally, in section~\ref{coherentstates} we describe the practical implementation technique of our quantum money scheme using coherent states and threshold detectors.  

\section{Materials and Methods}

This section focuses on the necessary definitions and tools required for the construction of our  quantum money scheme.

\subsection{Definitions for Private Quantum Money} \label{defin}

A private quantum money scheme involves an algorithm used by a trusted entity, Bank to produce multiple notes, and a  protocol that is run between a holder H of the note and the Bank to verify the authenticity of the note. The requirement for the verification protocol to be secure is that it must be impossible for an adversary noteholder to create more notes than what it received from the Bank.
\begin{defi}{Private quantum money.}
A quantum money scheme with classical verification consists of an algorithm by the Bank, and a  protocol, \emph{Verification}, such that,
\begin{enumerate}
\item Bank algorithm produces a quantum note $\$$ = ($\rho$, s.n.) where $\rho$ is a quantum state of the note and s.n. is the classical serial number of the note. 
\item \emph{Verification} is a protocol with classical communication that is run on the note $\$$, between the noteholder H who claims to possess the note $\$$ and the Bank. The output of the protocol is a bit declared by the Bank to denote whether the note is valid or not. We denote this final bit as $\text{Ver}^{B}_{H}(\$)$ which is 1 when the Bank validates the note and 0 otherwise.
\end{enumerate}
\end{defi}
For this scheme to be secure, it must satisfy two important properties,
\begin{itemize}
\item Correctness: The scheme is $\epsilon$ correct if for every honest holder H, it holds that

\begin{equation}
\mathbb{P}[\text{Ver}^{B}_{H}(\$) = 1] \geqslant 1 - \epsilon
\end{equation}
\item Unforgeability: The scheme is $\epsilon$ unforgeable if for any quantum adversary who possesses $m$ notes, has interacted a finitely bounded number of times with the Bank and has managed to produce $m'$ notes $\$_1,\$_2,\cdots,\$_{m'}$, it holds that,

\begin{equation}
\mathbb{P}\bigg[\bigg(\bigwedge_{i \in [m']}\text{Ver}^{B}_{H}(\$_i)=1\bigg)\wedge (m' > m)\bigg] \leqslant \epsilon
\end{equation}

\end{itemize}
where $H$ is any honest noteholder. The probability in the $\epsilon$-unforgeability property is taken over all possible strategies of an unbounded adversary.

The correctness condition ensures that all the honest noteholders get their note verified with an exponentially close to 1  probability (by setting $\epsilon$ exponentially close to 0). While the unforgeability condition ensures that an adversary trying to create more notes than what she had originally from the Bank, would fail with an exponentially close to 1 probability in being able to verify all the notes. Our definition includes the possibility of adaptive attacks by the adversary since we allow the interaction with the Bank a finite number of times during the verification protocol. 

{However, proving the security of such a general money-scheme can be a cumbersome task. To mitigate this  Aaronson and Christiano \cite{aaronson2012quantum} introduced the concept of a smaller \emph{public} money scheme (mini-scheme) and showed that it is sufficient to prove the security of this smaller version to guarantee security of the full scheme. Subsequently Ben-David and Sattath \cite{ben2016quantum} showed similar results for private money schemes using their construction of a private tokenized signature scheme. Such a signature scheme is used to produce signed documents that are publicly verifiable, unfeasible to forge by a third party, and has the property that the signing authority can produce and distribute one-use quantum signing tokens that allow the holder to sign only `one' document of her choice. The cryptographic assumption for the existence of such a private tokenized signature scheme is the existence of a collision-resistant hash function which is secure against quantum adversaries. There are various candidates for collision-resistant hash functions that are believed to be secure, so its reasonably valid assumption \cite{goldreich2004foundations}.} 

Under this mini-scheme, the Bank produces a single quantum note $\$$. The goal of the note adversary is, after finite interactions with the Bank, to produce two quantum notes $\$_1$ and $\$_2$ which successfully passes the verification test of the Bank. In this scheme, since the Bank produces only a single note $\$$, hence it does not require an attached classical serial number.
\begin{defi}{Private quantum money mini-scheme.}
A quantum money mini-scheme with classical verification consists of an algorithm by the Bank, and a protocol, \emph{Verification}, such that,
\begin{enumerate}
\item Bank algorithm produces a quantum note $\$$ = $\rho$ where $\rho$ is a quantum state of the note.
\item \emph{Verification} is a protocol with classical communication that is run on the note $\$$, between the noteholder H who claims to possess the note $\$$ and the Bank. The output of the protocol is a bit declared by the Bank to denote whether the note is valid or not. We denote this final bit as $\text{Ver}^{B}_{H}(\$)$ which is 1 when the Bank validates the note and 0 otherwise.
\end{enumerate}
\end{defi}
For this scheme to be secure, it must satisfy two important properties,
\begin{itemize}
\item Correctness: The scheme is $\epsilon$ correct if for every honest holder H, it holds that

\begin{equation}
\mathbb{P}[\text{Ver}^{B}_{H}(\$) = 1] \geqslant 1 - \epsilon
\end{equation}

\item Unforgeability: The scheme is $\epsilon$ unforgeable if for any quantum adversary who possesses the note $\$$, has interacted a finitely bounded number of times with the Bank and has managed to produce two notes $\$_1$ and $\$_2$, it holds that,

\begin{equation}
\mathbb{P}\bigg[\bigg(\text{Ver}^{B}_{H}(\$_1)=1 \wedge \text{Ver}^{B}_{H}(\$_1)=1 \bigg)\bigg] \leqslant \epsilon
\end{equation}

\end{itemize}
where H is any honest noteholder.

To go from a private quantum money mini-scheme to a full scheme, it is enough for the Bank to add a serial number to a note of the mini-scheme. Then the Bank can just run the verification protocol of the mini-scheme for that note with the serial number.
We, therefore, propose a quantum money mini-scheme and rely on the above results to extend this mini-scheme into full scheme.
\subsection{Tools for the Money Scheme} \label{tools}

In this sub-section, we describe the primary tool required for the construction of our money scheme, the Sampling Matching problem. This problem was originally defined by Kumar et al. to demonstrate a quantum advantage in one-way communication complexity setting \cite{kumar2019experimental}. We use a variant of the original problem to construct the verification scheme for the honest verifier.

\subsubsection{Sampling Matching problem} \label{SHM}

\begin{figure}[h!]
\includegraphics[scale=0.4]{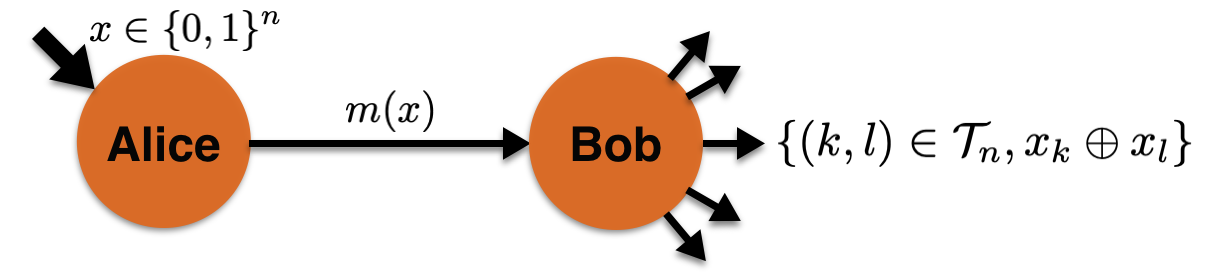}
\centering
\caption{Sampling matching problem. Alice receives an input $x \in \{0,1\}^n$  and Bob does not receive any input. Alice sends a message $m(x)$ to Bob who outputs the tuple $\{ (k,l) \in \mathcal{T}_n, x_k \oplus x_l\}$ where the from the message $m(x)$, a tuple $(k,l)$ is sampled from the set of possible distinct tuples $\mathcal{T}_n$. Bob's objective is to output the parity correctly with high probability.}
\label{fig:PHM1}
\end{figure}

The Sampling Matching problem as illustrated in Figure~\ref{fig:PHM1} consistes of two players, Alice and Bob.  For any positive integer $n$, Alice receives a binary string  $x \in \{0,1\}^n$. Bob, on the other hand, does not receive any input. {His task is to sample a tuple $(k,l)$ on the complete graph of $n$ vertices (with the vertices being indexed with numbers $\{1,2,\cdots, n\}$) uniformly at random from a set of $\mathcal{T}_n$ containing  $n(n-1)/2$ distinct tuples. An example of the tuple set for $n = 4$ is $\mathcal{T}_4$ : $\big[(1,2),(3,4),(1,3),(2,4),(1,4),(2,3) \big]$.} \par

The objective of the problem is for Bob to output any tuple $(k,l)$ from $\mathcal{T}_n$ and the parity $x_k \oplus x_l$ (where $x_k,x_l$ are the $k^{\text{th}}$ and $l^{\text{th}}$ bit of $x$ respectively).  We look at the model of one-way communication where we only allow a single message from Alice to Bob.

\begin{figure}[h!]
\includegraphics[scale=0.25]{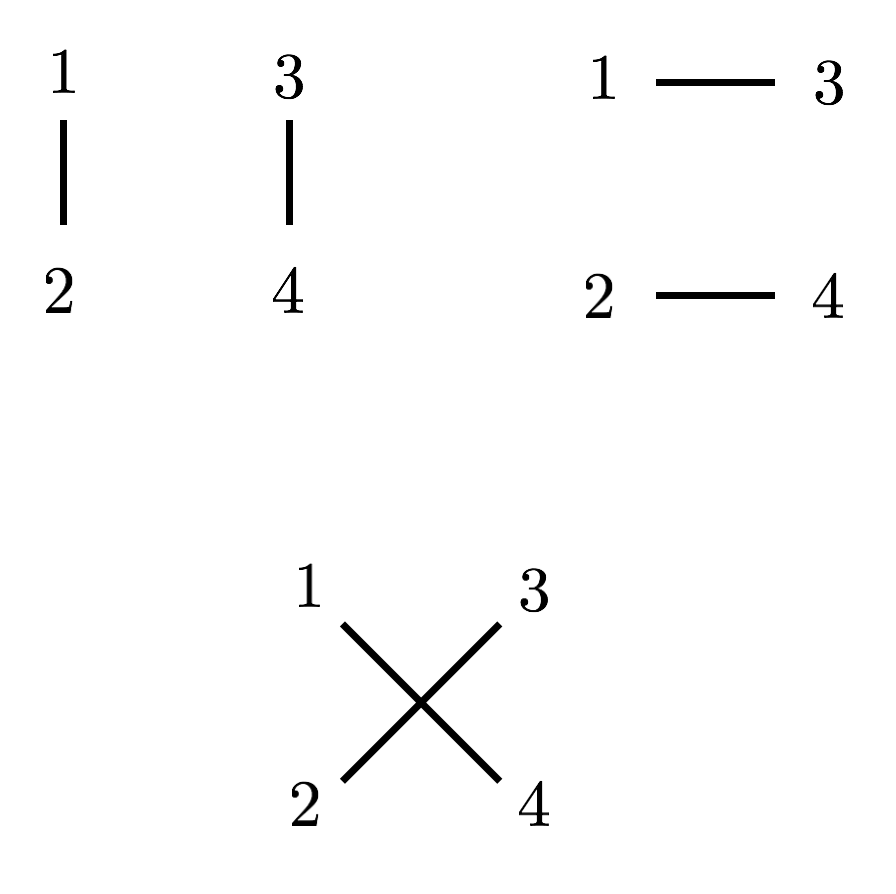}
\centering
\caption{An example of all possible distinct tuples for the size $n=4$: $\mathcal{T}_4$ : $\big[(1,2),(3,4),(1,3),(2,4),(1,4),(2,3) \big]$}
\label{fig:finger}
\end{figure}

For our quantum money proposal, we study the case when Alice is the untrusted noteholder and Bob is an honest note verifier. 
In the following, we construct a scheme for Bob to sample a tuple $(k,l)$ from the set $\mathcal{T}_n$ when Alice sends a quantum message to Bob. We analyse Bob's scheme when the message sent by Alice is a single-photon state in a superposition over $n$ modes.   

\subsubsection{Sampling Matching Scheme with Single-photon States} \label{SMSingle}

Sampling Matching scheme is Bob's testing scheme to sample the parity outcome of a tuple from the set $\mathcal{T}_n$ containing $n(n-1)/2$ distinct tuples. Here we look at the testing scheme when Alice's quantum message to Bob is a single-photon state.

\begin{figure}[h!]
\includegraphics[scale=0.36]{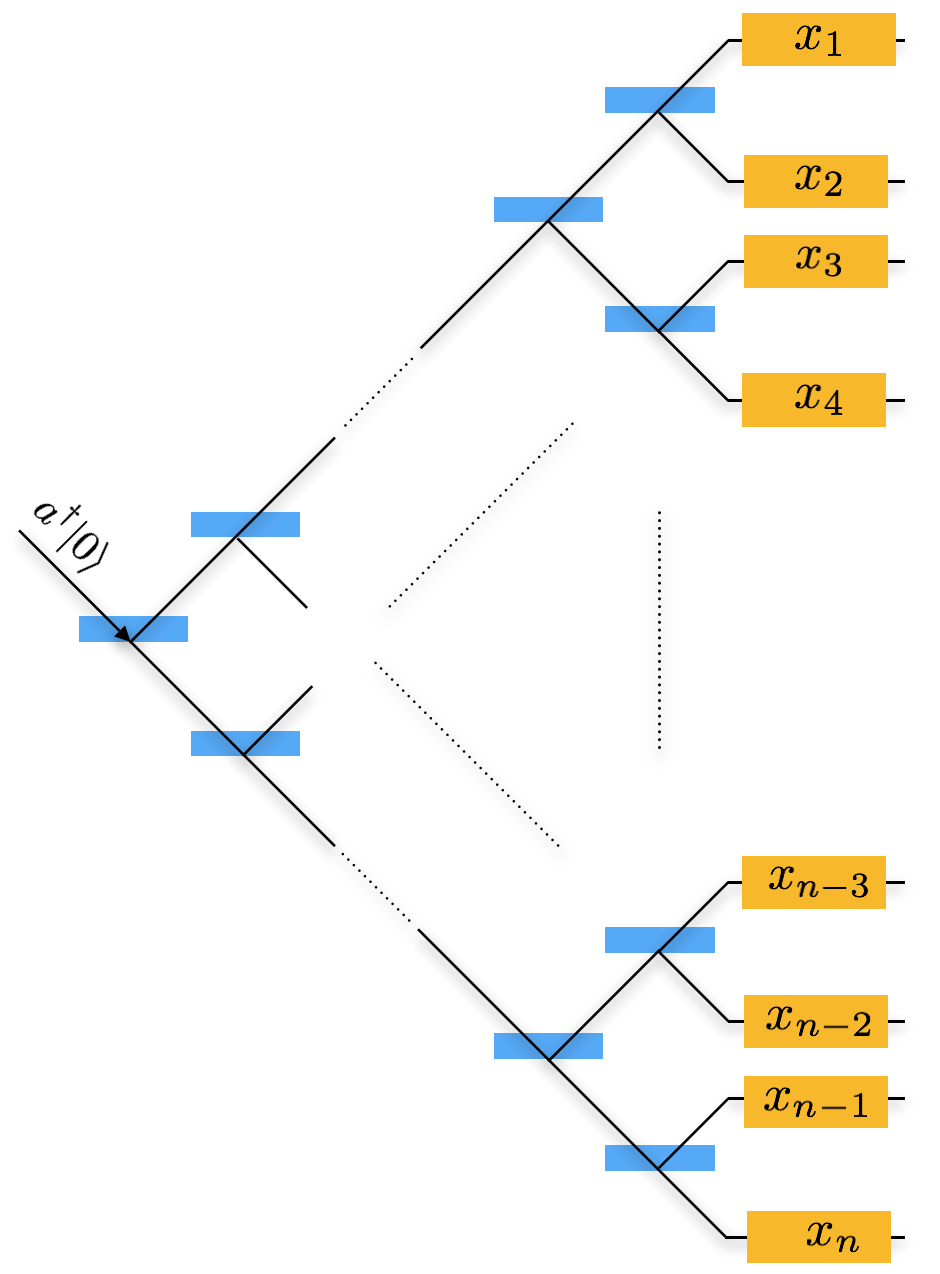}
\centering
\caption{Superposition circuit of Alice to create a single-photon state in equal superposition over $n$ modes. This is realised by passing a single-photon through a cascade of $n-1$ 50/50 beam splitters and subjecting each output mode to a phase-shift that depends on the corresponding  string $x \in \{0,1\}^n$.}
\label{Superposition}
\end{figure}

The technique is depicted as follows: When an honest Alice receives the binary string $x \in \{0,1\}^n$, she encodes the information of this string into a single-photon state in a superposition over $n$ different modes,

\begin{equation}
\ket{x} = \frac{1}{\sqrt{n}}\sum_{k=1}^{n}(-1)^{x_{k}}\hat{a}_k^{\dagger}\ket{0}, 
\label{Eq:12}
\end{equation} 

where $x_{k}$ is the $k$-th bit of the string $x$. The operator $\hat{a}_k^{\dagger}$ is the creation operator for the $k$-th mode, and, $\hat{a}_k^{\dagger}\ket{0} = \ket{1}_k$. Figure~\ref{Superposition} illustrates a method to create of equal superposition state of Eq.(\ref{Eq:12}) by passing the initial state $\hat{a}^{\dagger}\ket{0}$ through the cascade of $n-1$ 50/50 beam splitters and adding the phase information of each bit of the input in the $n$ modes. 

Alice sends this state $\ket{x}$ to Bob. In order to determine the parity outcome of a tuple, Bob first prepares his local superposition state in n-modes,

\begin{equation}
\ket{\beta} = \frac{1}{\sqrt{n}}\sum_{k=1}^{n}\hat{b}_k^{\dagger}\ket{0},
\end{equation}

where $\hat{b}_k^{\dagger}$ is the creation operator for the $k$-th mode with $\hat{b}_k^{\dagger}\ket{0} = \ket{1'}_k$. 

\begin{figure}[h!]
\includegraphics[scale=0.34]{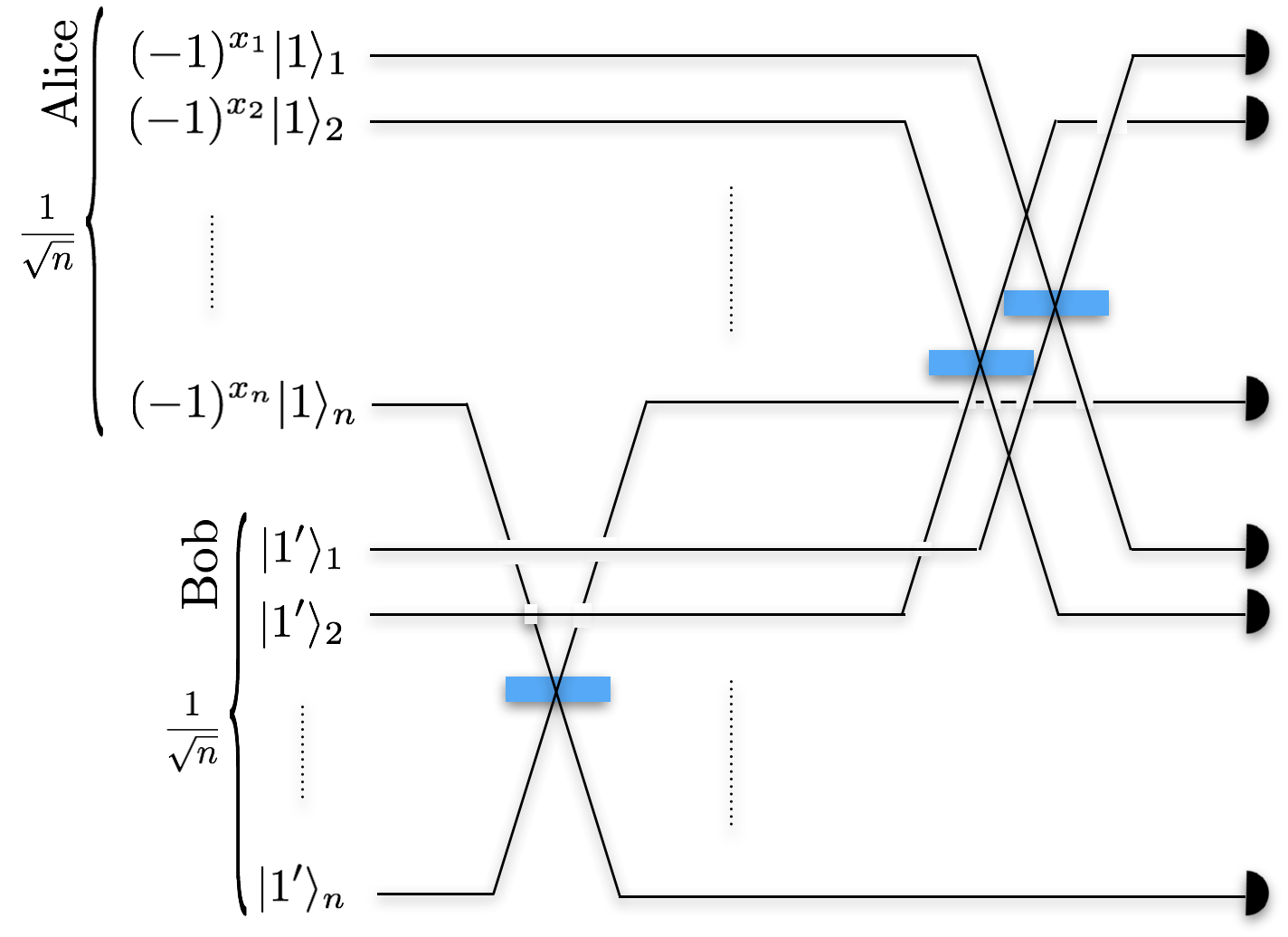}
\centering
\caption{Sampling Matching circuit model in single-photon encoding. Alice encodes a secret string $x \in \{0,1\}^n$ in the single-photon state $\ket{x}$ in an equal superposition over $n$ modes. This is then sent to Bob. Bob creates his local superposition state and applies mode-by-mode beam splitter operation with Alice's state. The results are observed in the $2n$ photon number resolving detectors.}
\label{SHMBS2}
\end{figure}

Bob's action is to apply mode-by-mode beam splitter operation on the state $\ket{x}\otimes\ket{\beta}$. This is illustrated in Figure~\ref{SHMBS2}.

Prior to the beam splitter operation, the input of Bob is,

\begin{equation}
\ket{x}\otimes\ket{\beta} = \frac{1}{\sqrt{n}}\sum_{k=1}^{n}(-1)^{x_{k}}\hat{a}_k^{\dagger}\ket{0}\otimes \frac{1}{\sqrt{n}}\sum_{l=1}^{n}\hat{b}_l^{\dagger}\ket{0} = \frac{1}{n}\sum_{k,l=1}^{n}(-1)^{x_k}\hat{a}_k^{\dagger}\hat{b}_l^{\dagger}\ket{00}
\end{equation}

where $\ket{00} = \ket{0}\otimes\ket{0}$.

\begin{figure}[h!]
\includegraphics[scale=0.25]{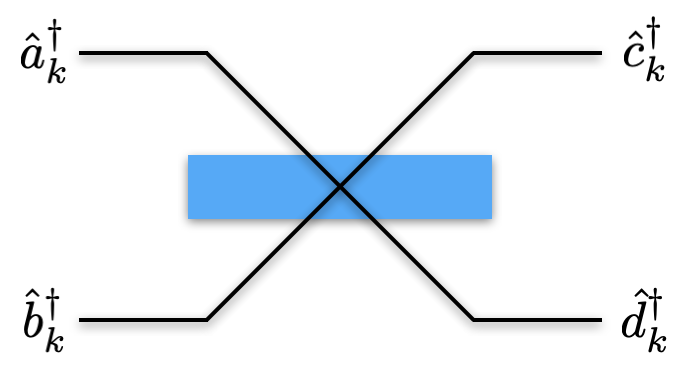}
\centering
\caption{Illustration of a 50/50 beam splitter transforming input modes $\{\hat{a}_k^{\dagger},\hat{b}_k^{\dagger}\}$ into the output modes $\{\hat{c}_k^{\dagger},\hat{d}_k^{\dagger}\}$.}
\label{BS1photon1}
\end{figure}

At each mode $k \in [n]$, the beam splitter transforms the input operators $\{\hat{a}_k^{\dagger},\hat{b}_k^{\dagger}\}$ into the output modes $\{\hat{c}_k^{\dagger},\hat{d}_k^{\dagger}\}$ as depicted in Figure~\ref{BS1photon1}. This input to output mode conversion for the 50/50 beam splitter is given as,
 
\begin{equation}
\hat{a}_k^{\dagger} \rightarrow \frac{1}{\sqrt{2}}(\hat{c}_k^{\dagger}+\hat{d}_k^{\dagger}), \hspace{2mm} \hat{b}_k^{\dagger} \rightarrow \frac{1}{\sqrt{2}}(\hat{c}_k^{\dagger}-\hat{d}_k^{\dagger})
\end{equation} 

{From Figure~\ref{SHMBS2}, we see that the $k$-th mode of Alice's state interacts with $k$-th mdoe of Bob's state, for all $k \in [n]$. The output operator $\hat{O}^{\dagger}$ corresponding to this joint interaction for all the modes is,}
{
\begin{equation}
\begin{split}
\hat{O}^{\dagger}&= \frac{1}{n}\sum_{k,l=1}^{n}(-1)^{x_{k}}\frac{\hat{c}_k^{\dagger}+\hat{d}_k^{\dagger}}{\sqrt{2}}\cdot \frac{\hat{c}_l^{\dagger}-\hat{d}_l^{\dagger}}{\sqrt{2}}, \\
&= \frac{1}{2n}\sum_{k,l=1}^{n}(-1)^{x_{k}}(\hat{c}_k^{\dagger}\hat{c}_l^{\dagger} + \hat{d}_k^{\dagger}\hat{c}_l^{\dagger} -\hat{c}_k^{\dagger}\hat{d}_l^{\dagger} - \hat{d}_k^{\dagger}\hat{d}_l^{\dagger}), \\
&= \frac{1}{2n}\sum_{k=1}^{n}(-1)^{x_{k}}(\hat{c}_k^{\dagger 2} - \hat{d}_k^{\dagger 2}) \hspace{2mm} + \\
&\hspace{5mm} \frac{1}{2n}\sum_{(k,l)\in \mathcal{T}_n}((-1)^{x_{k}} + (-1)^{x_{l}})(\hat{c}_k^{\dagger}\hat{c}_l^{\dagger} - \hat{d}_k^{\dagger}\hat{d}_l^{\dagger}) \hspace{2mm} + \\
&\hspace{5mm}  \frac{1}{2n}\sum_{(k,l)\in \mathcal{T}_n}((-1)^{x_{k}} - (-1)^{x_{l}})(\hat{d}_k^{\dagger}\hat{c}_l^{\dagger} - \hat{c}_k^{\dagger}\hat{d}_l^{\dagger})
\end{split}
\label{Eq:C4}
\end{equation}}
{where the commutation $[\hat{c}_k^{\dagger}, \hat{d}_l^{\dagger}] = 0$ for all $k,l \in [n]$. This is because by notation $\hat{c}^{\dagger}$ acts on the first qubit while the $\hat{d}^{\dagger}$ acts on the second qubit. Here $\mathcal{T}_n$ is the set of all possible $\frac{n(n-1)}{2}$ distinct tuples.}
The output state is,
\begin{equation}
\hat{O}^{\dagger}\ket{00}
\label{Eq:Output}
\end{equation}

{The output Eq.(\ref{Eq:Output}) is then observed in the photon number resolving detectors. Since there are two input photons to the circuit, Bob will observe two-photons clicks across the $2n$ output modes (here we assume perfect set-up and detection) labelled as $ \{c_1,d_1,\cdots,c_n,d_n\}$. From Eq.(\ref{Eq:C4}) Bob will observe one of the following scenarios,}
\begin{itemize}\label{simclicks}
\item Simultaneous single-photon clicks in $\{{c}_k,{c}_l\}$ or $\{{d}_k,{d}_l\}$, for two distinct modes $(k,l)$, implies $x_{k} \oplus x_{l} = 0$.
\item Simultaneous single-photon clicks in $\{{c}_k,{d}_l\}$ or $\{{d}_k{c}_l\}$, for two distinct modes $(k,l)$, implies $x_{k} \oplus x_{l} = 1$.
\item 2 photons in the same mode ${c}_k$ or ${d}_k$ does not reveal the parity outcome for Bob and hence results in inconclusive outcome.
\end{itemize}
{We denote the output state corresponding to single-photon detection in the two distinct modes $\{c_k,d_l\}$ as $\ket{11}_{c_k,d_l}$. The same notation is followed for output states corresponding to single-photon detection in distinct modes $\{c_k,c_l\}$ or $\{d_k,d_l\}$. The output state corresponding to two-photon detection in the same modes $c_k$ or $d_k$ is denoted as $\ket{20}_{c_k}$ and $\ket{02}_{d_k}$ respectively.} 

From Eq.(\ref{Eq:C4}), we see that the probability of observing 2 photons in the same mode ${c}_k$ is, 
\begin{equation}
p_{2}^{{c}_k} = |\bra{20}_{c_k}\hat{O}^{\dagger}\ket{00}|^2 = \frac{1}{2n^2}
\end{equation}
where we have used the property of creation operators that $\hat{c}_k^{2\dagger}\ket{00} = \sqrt{2}\ket{20}_{c_k}$.
Similarly, the probability of observing 2 photons in the same mode $\hat{d}_k$ is,
\begin{equation}
p_{2}^{{d}_k} = |\bra{02}_{d_k}\hat{O}^{\dagger}\ket{00}|^2 = \frac{1}{2n^2}
\end{equation}
where the similar property of creation operators $\hat{d}_k^{2}\ket{00} = \sqrt{2}\ket{02}_{d_k}$ has been used. Over all the $2n$ modes, the probability of having 2 photons in the same mode is,
\begin{equation}
p_2 = \sum_{k=1}^{n}p_{2}^{{c}_k} + p_{2}^{{d}_k} = \frac{1}{n}
\label{2photonsamemode}
\end{equation}
In these cases, Bob does not get a conclusive parity outcome of any two bits of $x$. When this occurs, he outputs the outcome $d = \varnothing$.

 On the other hand, with probability $1 - \frac{1}{n}$, Bob always gets exactly two single-photon clicks in two different time modes $k,l \in [n]$ with correct parity outcome $d = x_k \oplus x_l$. The probability that he outputs the parity outcome of a tuple $(k,l) \in \mathcal{T}_n$,

\begin{equation}
p_{kl} = |\bra{00}\hat{T}_{kl} \cdot O^{\dagger}\ket{00}|^2 = \frac{2}{n^2}
\label{eq:C5}
\end{equation}

where $\hat{T}_{kl}^{\dagger} = \frac{1}{2\sqrt{2}}\bigg(((-1)^{x_{k}} + (-1)^{x_{l}})(\hat{c}_k^{\dagger}\hat{c}_l^{\dagger} - \hat{d}_k^{\dagger}\hat{d}_l^{\dagger}) + ((-1)^{x_{k}} - (-1)^{x_{l}})(\hat{c}_k^{\dagger}\hat{d}_l^{\dagger} - \hat{d}_k^{\dagger}\hat{c}_l^{\dagger})\bigg)$ is the operator corresponding to the correct parity outcome for the tuple $(k,l)$. Operationally this operator generates only single-photon clicks in modes $\{c_k,c_l$ or $\{d_k, d_l\}$ if $x_k \oplus x_l = 0$, while single-photon clicks are generated in $\{c_k,d_l$ or $\{d_k, c_l\}$ if the corresponding parity is 1. {Note that for the incoming state $\ket{x}$ from honest Alice, if Bob observes simultaneous single-photon clicks in two distinct modes, it's parity outcome is correct with certainty.} Hence if Alice is dishonest and sends a state different from $\ket{x}$, it is enough to show that after the mode-by-mode beam splitter interaction with Bob's state $\ket{\beta}$, there is a non-zero probability for Bob to obtain an incorrect parity outcome $x_k \oplus x_l$ across the tuples $(k,l) \in \mathcal{T}_n$. This forms the basis of our unforgeability test using Sampling Matching based verification.  

\section{Results}

With the tools at our disposal, we can now formally introduce our private quantum money scheme using the verification protocol based on the Sampling Matching scheme.

\subsection{Private Quantum Money Scheme}  \label{QuantumMoney}

{Our quantum money scheme involves a Bank that produces the notes which are quantum states and distributes them to untrusted noteholders. Each time a holder needs to carry out a transaction, he sends the note to the trusted local verifiers of the Bank who help the Bank decide the validity of the note.}  The salient features of our money scheme include:
\begin{itemize}
\item Note verification procedure requiring a single round classical communication between the local verifier and the Bank,
\item Fixed verification circuit for a given input size of the note,
\item Multiple note re-usability, meaning the same note can be reused by the holder a number (linear in the size of the note) of times,
\item Unconditional security against any adversary trying forge the banknote while tolerating a noise of up to $21.4\%$. 
\end{itemize} 

Our scheme involves two phases. First is the \emph{Note preparation phase}, where the Bank chooses multiple n-bit binary strings independently and uniformly randomly. Each of these strings are then encoded into single-photon states in superposition over $n$ modes. The quantum note $\$$ is the combination of single-photon states corresponding to all the chosen  strings. This is then distributed among the untrusted holders. In the \emph{Verification phase}, the holder's note is sent to the local verifier for testing. The verifier randomly selects some copies of the note state (here the note consists of multiple copies, where one copy corresponds to the single-photon state that encodes one n-bit string). He runs the verification protocol using the Sampling Matching, SM scheme (Section~\ref{SMSingle}) and does a local check based on the statistics of the classical outcome obtained. If the statistics are different from the ones produced by the note of an honest holder, he invalidates the note. If the note passes this test then the outcomes are classically communicated to the Bank. The Bank compares these outcomes with his private $n$-bit strings. If a high fraction of the outcomes is correct, he outputs the bit $\text{Ver}^{B}_{H} =1$ implying that the note is valid. Otherwise, he outputs $\text{Ver}^{B}_{H} = 0$.  

The money scheme we use here is the quantum money mini-scheme. Under this scheme, the Bank produces a single quantum note $\$$, consisting of many copies of single-photon states. The goal of the note adversary is, after finite interactions with the Bank, to produce two quantum notes $\$_1$ and $\$_2$ which successfully pass the verification test by two independent verifiers. {We have already emphasized that we make use of previous results \cite{ben2016quantum} which lifts the security against any adversary from a private quantum money mini-scheme to the general scheme with multiple notes and a classical serial number assigned to them.}

We now describe the quantum money mini-scheme based on single-photon states, 50/50 beam splitter linear optics transformations and photon number resolving detectors. 

\begin{figure}[h!]
\includegraphics[scale=0.52]{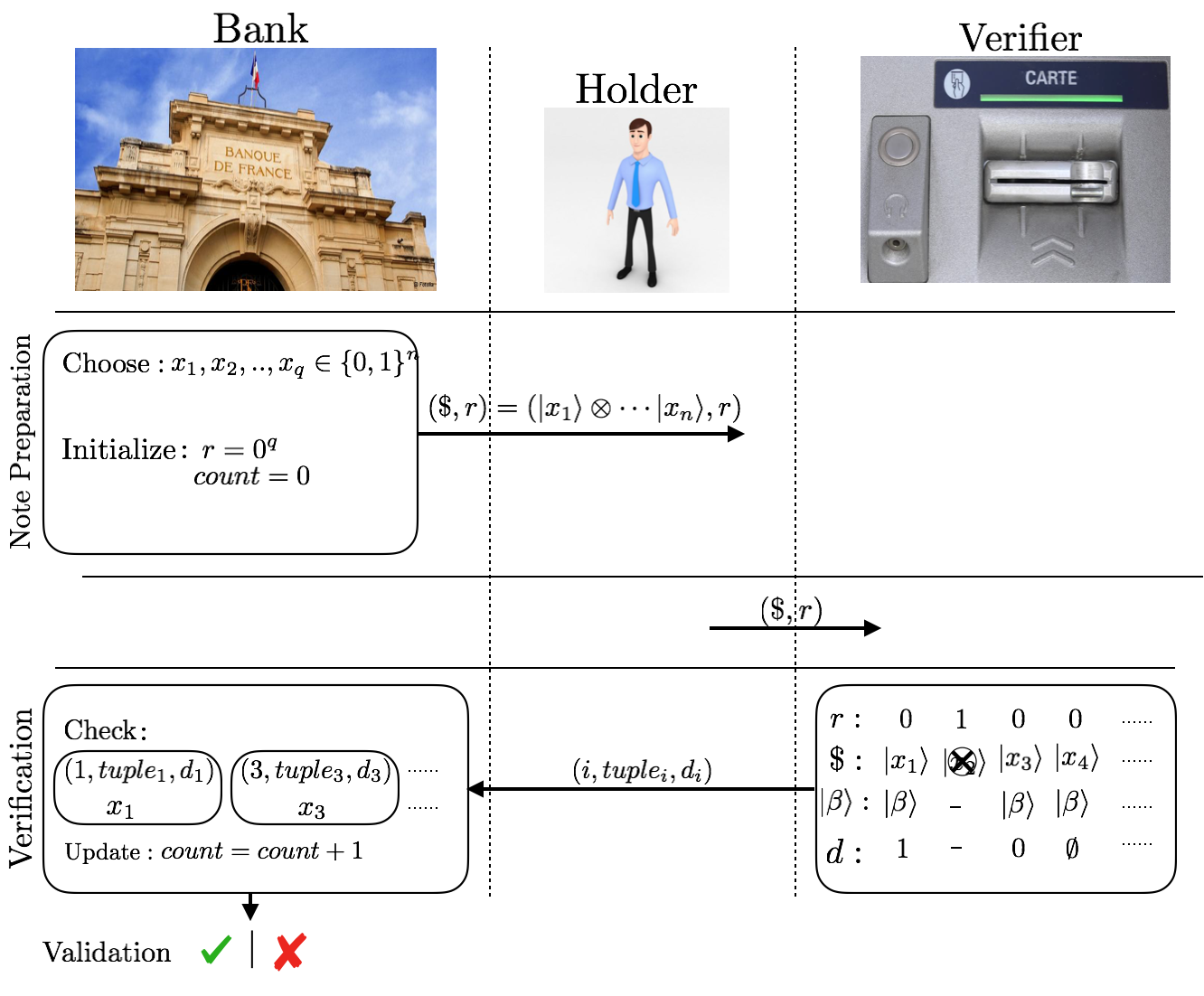}
\centering
\caption{Illustration of our private quantum money scheme based on the verification protocol using the Sampling Matching scheme. In the \emph{Note Preparation} phase, the Bank independently and uniformly randomly selects $q$ n-bit binary strings $\{x_1,\cdots,x_q\}$ and prepares single-photon superposition note states $\$ = \ket{x_1}\otimes\ket{x_2}\otimes\cdots\otimes\ket{x_q}$. The Bank further initializes a $r$ register to keep a track of the number of positions in $[q]$ where the states have been used for verification and the $count$ register to keep track of a number of verification attempts on the note. The note $(\$, r)$ is then sent to the holder. To be able to carry out any transaction, the holder sends the note to an honest verifier. In the \emph{Verification} phase, the verifier selects a fraction of the $q$ copies of the note state which have an $r = 0$. He creates his local state $\ket{\beta}$ and applies the SM-scheme on those selected copies. The verifier sends the outcome of the measurement scheme to the Bank. Finally, the Bank compares the outcomes with his secret string $x_j$'s  and outputs a bit $\text{Ver}^{B}_{H}$ stating whether the note is valid or not.}
\label{fig:SHMPicSP}
\end{figure}

\subsubsection{Note Preparation Phase}

\begin{enumerate}
  \item The Bank independently and uniformly randomly chooses $q$ $n$-bit binary strings $x_1,x_2,..,x_q \in \{0,1\}^n$ 
  \item The Bank encodes each binary string $x_j$ into the single-photon state in superposition over $n$ modes,
  \begin{equation}
  \ket{x_j} = \frac{1}{\sqrt{n}}\sum_{k=1}^{n} (-1)^{x_{j,k}}\hat{a}_k^{\dagger}\ket{0}
  \label{Eq:HonestNote}
  \end{equation}
  where $x_{j,k}$ is the $k$-th bit value of string $x_j$ and $\hat{a}_k^{\dagger}$ is the creation operator  for the mode $k$ with $\hat{a}_k^{\dagger}\ket{0} = \ket{1}_k$. {The note is a combination of all the $q$ copies of the single-photon state $\$ = \ket{x_1}\otimes\ket{x_2}\otimes\cdots\otimes\ket{x_q}$.}
  \item The Bank creates a classical binary register $r$ and initializes it to $0^q$. This register keeps  the track of positions $j$ where the states  have been used for the verification.
  \item The Bank  creates a counter variable $count$ and initializes it to 0. This keeps a track of the number of verification attempts. 
  \item  The Bank sends the quantum note ($\$$, r) to the holder.\\
\end{enumerate} 

\subsubsection{Verification Phase}

Once the Bank distributes the note, the holder in order to carry out any transaction has to get the note verified from an honest local verifier Ver. The verification procedure is described below. 
\\\\
\textbf{Local testing}
\begin{enumerate}
\item The holder gives the note $\$' (=: \$$ if the holder is honest) to Ver. 
\item Ver checks for the number of times the note has be re-used by verifying that the hamming distance of $r$ register $d(r,0^q) \leqslant T$, where $T$ is a predefined maximum number of copies in the note that are allowed for verification. If $d(r,0^q) > T$, the note is rendered useless.
\item Ver uniformly and randomly selects a subset $L \subset [q]$ copies from the states marked $r = 0$. All the corresponding $L$ copies in the $r$ register is then marked to 1.
\item For each chosen copy $j \in L$, Ver prepares his local coherent state $\ket{\beta}$ and runs the SM scheme (Section~\ref{SMSingle}). 
\item {Ver first checks if he gets 2 photon clicks in all the chosen $L$ copies. If not, he rejects (this is a check against all those attacks where the adversary removes the single-photon state and introduces either vacuum or a multi-photon state).}
\item Ver counts the number of successful copies $l_{succ}$, where he obtains two single-photon clicks in two different modes. For these copies he outputs the parity outcome  $d_j = x_{j,k} \oplus x_{j,l}$ where the clicks have been obtained in modes $k$ and $l$. For the rest of the copies, he sets $d_j = \varnothing$.
\item Ver checks if $l_{succ} \geqslant l_{min}$, where $l_{min} = \mathbb{E}_{h}[l_{succ}](1 -\epsilon)$ is the  minimum number of copies that will locally guarantee his acceptance of the note, where $0\leqslant\epsilon\leqslant 1$ is the desired security factor. Here $\mathbb{E}_{h}[l_{succ}]$ is the expected number of copies where the honest noteholder obtains two single-photon clicks in two different modes when Ver runs the SM scheme.
\item Ver proceeds to the classical communication step with the Bank only when the note passes this test.
\end{enumerate}
\textbf{Communication with the Bank}
\begin{enumerate}
\item {Ver forwards the outcomes $\{j,(k,l),d_j\}$ for each $j \in L$ to the Bank through a classical authenticated channel}.
\item The Bank checks if $count < \lceil \frac{T}{|L|} \rceil$, otherwise the verification attempt is rendered invalid. Here $\lceil \cdot \rceil$ is the ceiling function. 
\item For each copy $j \in L$ with $d_j \neq \emptyset$, the Bank compares the parity value $d_j$ with the secret string $x_j$. He validates the note if the number of correct outcomes
\begin{equation}
l_{succ}^{cor} \geqslant \mathbb{E}_h[l_{succ}^{cor}](1 - \delta)  
\end{equation}
where $\mathbb{E}_h[l_{succ}^{cor}]$ is the expected number of copies that give the correct parity outcome when the noteholder is honest, and $0\leqslant\delta\leqslant 1$ is a positive constant whose optimal value is determined by the forging probability (Section~\ref{unfor}).
\item The Bank updates the $count$ by 1.
\end{enumerate}

\subsection{Correctness} \label{HonestHolder}

Let us start by computing the probability that an honest noteholder fails the verification test. We use the Chernoff-Hoeffding inequality \cite{upfal2005probability} to prove our results.  

We first remark that the honest noteholder always passes the step 5 of the \emph{Verification phase}, since he sends the entire banknote to the verifier Ver, who after performing the SM-scheme on the chosen $L$ copies, always obtains the two-photon clicks. 

However, the noteholder can fail the step 7 of the \emph{Verification phase} if the number of successful copies, where he obtains two single-photon clicks in two different time modes, $l_{succ} < l_{min} = \mathbb{E}[l_{succ}](1 - \epsilon)$, where $\mathbb{E}[l_{succ}]$ is the expected number of copies where Ver obtains two single-photon clicks in two different modes when he runs the SM scheme, and $\epsilon$ is the security parameter chosen by Ver. Eq.(\ref{2photonsamemode}) tells us that for each of these chosen copy $j \in L$, the probability that the verifier obtains two single clicks in two different time modes is,

\begin{equation}
p_{11} = 1 - \frac{1}{n}
\end{equation}

Thus for $L$ copies chosen from the note state, the expected number of successful copies is, 

\begin{equation}
\mathbb{E}[l_{succ}] = |L|
p_{11}
\label{Eq:HLSUCC}
\end{equation}

Using the Chernoff-Hoeffding bound, the probability that the holder fails this test is,

\begin{equation}
\mathbb{P}[l_{succ} < l_{min}] \leqslant \exp\bigg(-\frac{2\epsilon^2\mathbb{E}^2[l_{succ}]}{|L|}\bigg) = \exp(-2\epsilon^2p_{11}^2|L|)
\label{Eq:honestlsucclesslmin}
\end{equation}

Now after applying the SM scheme, Ver forwards the parity outcomes to the Bank. From Eq.(\ref{Eq:C4}), we see that whenever Ver obtains a parity outcome of a tuple $(k,l)$, it is always correct. Thus the probability that the Bank obtains correct parity outcome for each of the  $l_{succ}$ copy is $c = 1$. This implies $l_{succ}^{cor} = l_{succ}$ and that the only stage where an honest noteholder can fail is the local verifier stage.

\begin{equation}
\mathbb{P}[\text{Honest fail}] = \mathbb{P}[l_{succ} < l_{min}]\leqslant  \exp(-2\epsilon^2(1 - \frac{1}{n})^2|L|)
\label{Eq:honestprefect}
\end{equation}

This probability of failing goes down exponentially with $\epsilon, |L|$ and $n$.

\subsection{Unforgeability of Banknotes} \label{unfor}

In this section, we explicitly calculate the forging probability for the adversary when she has in possession the valid banknote $\$$ and his objective is to duplicate the note to create two copies $\$_1$ and $\$_2$, which successfully pass the verification tests from two independent verifiers, Ver1 and Ver2, simultaneously\footnote{{Our proof technique has been inspired from the work of Amiri et al. \cite{amiri2017quantum} which is a private quantum money scheme with the verification based on Hidden Matching \cite{bar2004exponential}, another well-defined problem in one-way communication complexity. Kumar et al. \cite{kumar2019experimental} showed similarities between Sampling Matching and Hidden Matching communication problems. Here we manage to find that our proof of unforgeability has a similar structure to \cite{amiri2017quantum} even though the circuit to realise them is vastly different.}} \par

{Based on our scheme construction, any action by the adversary who is trying to maximize the forging probability can be put into three categories as specified detailed below. The first two categories, register manipulation, and adaptive forging, arise due to the addition of multiple reusability feature in our note. This can be leveraged by the adversary by performing `auxiliary' verification attempts on the note and thereby learning more information on the note state leading to an increase in the forging probability. The third category relates to an optimal quantum operation performed on the note state such that the resulting states sent to Ver1 and Ver2 can pass the verification test with high enough probability. Our proof of unforgeability takes into account the optimal action by any adversary on all these three categories.}

\textbf{1. Register manipulation}. First, we address how manipulating the $r$ register can be used by an adversary to increase the forging probability.  Since in each verification attempt the verifier chooses $|L|$ copies from the note state, and the maximum number of note verification attempts allowed by the Bank is $T$, hence the adversary can set at most $(T-1)|L|$ positions in the $r$ register to 1 before sending the note state to the verifier. {This ensures that the verifier does not render the note useless while checking the hamming distance of the $r$ register. Further, since the verifier only selects copies from the note state marked 0 in the $r$ register, setting maximum possible positions in $r$ to 1 allows the adversary the need to deal with lesser copies (0 marked states) of the note state.} 

Suppose the adversary creates two notes $(\$_1,r_1)$ and $(\$_2,r_2)$  and sends it to the verifiers Ver1 and Ver2 respectively. If the adversary sets $r_1(j) = 0$ and $r_2(j) = 1$ for the $j$-th copy of the note, she is sure that Ver2 will not select the state at $j$-th position for verification. With this approach she can set a maximum of $(T-1)|L|$ positions in $r_1$ and $r_2$ register to 1. In the positions where she has set $r_1 = 1$, she can send the correct banknote states to Ver2, and similarly for the positions where she has set $r_2 = 1$, she can send banknote states to Ver1. This results in her  exactly replicating the $2(T-1)|L|$ copies of the note state for both verifiers. 
 
\textbf{2. Adaptive forging}. Let us now also consider the possibility of the adaptive attack by the adversary where multiple `auxiliary' verification attempts made on the note state would help increase her forging probability. Since the Bank allows a maximum of $T$ verification attempts on the note, hence the adversary can use this to his advantage by querying for validation of $(T-2)|L|$ copies (since he needs to leave 2 verification attempts, one each from Ver1 and Ver2). We assume the worst-case scenario where the adversary completely obtains the information of the state for those copies used in auxiliary verification attempts. \par

{The above two categories of attacks ($r$ register manipulation and the adaptive attack) combined allow the adversary to have full information of a combined $(3T-4)|L|$ copies of the note state. }

\textbf{3. General forging}. Now to prove the unforgeability condition, we consider what happens in the remaining $q' = q - (3T-4)|L|$ copies of the states sent to Ver1 and Ver2 where the adversary has no auxiliary information of the states and for which $r_1(j)$ and $r_2(j)$ are 0. {In this scenario, an optimal adversary operation is to produce two note states with $q'$ copies each (one each for Ver1 and Ver2) such that the average fidelity of the prepared states with the correct note state is maximised. This maximisation directly translates to the probability with which verifier obtains correct outcomes upon performing the Sampling Matching scheme. \footnote{{The problem of fidelity estimation can be cast directly as a semi-definite optimisation (SDP) problem to upper bound the adversary's forging probability \cite{molina2012optimal}.}} There are two possible attack models that the adversary can employ on the note state. First is the \textbf{collective} attack based on optimal manipulation on individual copies of the note state. The second, more general model, is \textbf{coherent} attack where the adversary performs a combined operation on the entire note state. We start with the collective attack and subsequently use the results by Croke and Kent \cite{croke2012security} to argue that any coherent strategy by the adversary is no more powerful than the optimal collective strategy.} \par

\textbf{3.1 Collective attack}: Here we look at the optimal manipulation strategy on individual copies of the note state. First, we remark that the adversary has to send a single-photon state across each the $q'$ copies to Ver1 and Ver2, otherwise he fails the step 5 test in \emph{Verification phase} with certainty.
For each copy $j \in [q']$, the adversary possesses the valid banknote state $\ket{x_j}$ ( Eq~\ref{Eq:HonestNote}). Using this copy, his most general operation would be,
{
\begin{equation}
    \ket{x_j}\otimes\ket{\text{anc}} \xrightarrow{\varepsilon} \rho_{j,HV_1V_2}
\end{equation} }
{where $\rho_{j,HV_1V_2}$ is a general tri-partite quantum state that the adversary (H) creates between her, Ver1 ($V_1$) and Ver2 ($V_2$). The state received by Ver1 is then $\eta_{x_j} = Tr_{HV_2}(\rho_{j,HV_1V_2})$, where $Tr_{HV_2}(.)$ is partial trace over adversary and Ver2's state. Similarly the state received by Ver2 is $\tau_{x_j} = Tr_{HV_1}(\rho_{j,HV_1V_2})$.} Any normalized mixed state sent to Ver1 over $n$-modes can be expressed as,
\begin{equation}
\eta_{x_j} = \sum_{k,l}^{n}A_{kl}\hat{a}_k^{\dagger}\ket{0}\bra{0}\hat{a}_l
\label{Eq:D2}
\end{equation} 

where $\hat{a}_k^{\dagger}$ is the creation operator of the $k^{th}$ mode, and $A_{kl} 
\in \mathbb{C}$ for all $k,l \in [n]$. The normalization of the mixed state invokes the condition $\sum_{k=1}^{n}A_{kk} = 1$. 

Ver1 runs the Sampling Matching scheme on the input state as shown in Figure~\ref{SHMBS2}.
The input of the interaction of the adversary state with the local state of Ver1, $\ket{\beta} = \frac{1}{\sqrt{n}}\sum_{k=1}^{n}\hat{b}_k^{\dagger}\ket{0}$, can be written as a combined density matrix state,

\begin{equation}
\begin{split}
\eta_{j}^{In} &= \ket{\beta}\otimes \eta_{x_j}\otimes \bra{\beta} = \frac{1}{n}\bigg(\sum_{k=1}^{n}\hat{b}_k^{\dagger} \ket{0}\otimes\sum_{l,m}^{n}A_{lm}\hat{a}_l^{\dagger}\ket{0}\bra{0}\hat{a}_m\otimes \sum_{o=1}^{n}\bra{0}\hat{b}_o\bigg) 
\label{eq:D3}
\end{split}
\end{equation}

This state undergoes mode-by-mode beam splitter operation (Figure~\ref{SHMBS2}) resulting the transformation of the modes $\hat{a}_i \rightarrow \frac{\hat{c}_i + \hat{d}_i}{\sqrt{2}}$ and $\hat{b}_i \rightarrow \frac{\hat{c}_i - \hat{d}_i}{\sqrt{2}}$ for all $i \in [n]$. The corresponding output state is,

\begin{equation}
\begin{split}
\eta_{j}^{Out} &= \frac{1}{4n}\bigg(\sum_{k=1}^{n}(\hat{c}_k^{\dagger} - \hat{d}_k^{\dagger}) \ket{0}\hspace{2mm} \otimes  \sum_{l,m=1}^{n}A_{lm}(\hat{c}_l^{\dagger} + \hat{d}_l^{\dagger})\ket{0}\bra{0}(\hat{c}_m + \hat{d}_m) \hspace{2mm} \otimes  \sum_{o=1}^{n}\bra{0}(\hat{c}_o - \hat{d}_o)\bigg) \\
&= \frac{1}{4n}\bigg(\sum_{k,l,m,o=1}^{n}A_{lm}(\hat{c}_k^{\dagger} - \hat{d}_k^{\dagger})(\hat{c}_l^{\dagger} + \hat{d}_l^{\dagger})\ket{00}\bra{00}(\hat{c}_m + \hat{d}_m)(\hat{c}_o - \hat{d}_o)\bigg)
\end{split}
\label{eq:mixedout}
\end{equation}

Similar to the analysis in section~\ref{SMSingle}, we first compute the probability ($p_{11}$) with which Ver1 observes single-photon clicks in two distinct modes. This probability $p_{11} = 1 - p_{2}$, where $p_2$ is total probability of observing two-photon clicks in the same mode. For output mode $c_k$,

\begin{equation}
p_{2}^{{c}_k} = \bra{20}_{c_k}\eta_{j}^{Out} \ket{20}_{c_k} = \frac{A_{kk}}{2n}
\end{equation}

The corresponding probability of observing two-photon clicks in mode ${d}_k$ is,

\begin{equation}
p_{2}^{{d}_k} = |\bra{02}_{d_k} \eta_{j}^{Out}\ket{02}_{d_k} = \frac{A_{kk}}{2n}
\end{equation}

Summing over all the $2n$ modes,

\begin{equation}
p_2 = \sum_{k=1}^{n}p_{2}^{{c}_k} + p_{2}^{{d}_k} = \sum_{k=1}^{n}\frac{A_{kk}}{n} = \frac{1}{n}
\label{2photonsamemodeAdv}
\end{equation}

where we have used the normalisation condition $\sum_{k=1}^{n}A_{kk} = 1$. Comparing Eq.(\ref{2photonsamemode}) and Eq.(\ref{2photonsamemodeAdv}), we see that the total probability of obtaining 2 photons in the same mode for an adversary is the same as that for an honest noteholder. Thus for any adversarial state with a single-photon over $n$ modes, Ver1 observes two single-photon clicks in two distinct output modes with a probability $p_{11} = 1 - \frac{1}{n}$. Over the $L$ copies chosen by Ver1, he receives on average $|L|p_{11}$ copies with single-photon clicks in two distinct modes. This implies that the adversary passes the local \textbf{step 7} of Ver1's \emph{Verification Phase} test with probability,

\begin{equation}
\mathbb{P}[\text{Ver1 accepts}] = \mathbb{P}[l_{succ} \geqslant l_{min}] \geqslant 1 - \exp(-2\epsilon^2p_{11}^2|L|)
\end{equation}

where $l_{succ}$ is the total copies where he gets single-photon clicks in two different modes and $l_{min} = |L|p_{11}(1 - \epsilon)$. If the adversary passes this test then Ver1 communicates the parity outcomes of $l_{succ}$ copies to the Bank who checks the outcomes with his secret strings. 

{We now calculate the average probability that the tuple $(e,f) \in \mathcal{T}_n$ that Ver1 obtains for the $j$-th copy, returns an incorrect parity outcome. Based on the analysis in section~\ref{SMSingle}, we see that an incorrect  outcome is obtained if Ver1 obtains single-photon clicks in the modes $\{c_e,c_f\}$ or $\{d_e,d_f\}$ when $x_{j,e} \oplus x_{j,f} = 1$, while single-photon clicks are obtained in $\{c_e,d_f\}$ or $\{d_e,c_f\}$ if the corresponding parity is 0. This probability of obtaining an incorrect parity outcome for the tuple $(e,f) \in \mathcal{T}_n$ is,}

\begin{equation}
p_{Ver1}^{ef,inc} = \bra{00}\hat{I}_{ef} \cdot \eta_{j}^{Out} \cdot \hat{I}_{ef}^{\dagger} \ket{00}
\label{Eq:D6}
\end{equation}

where $\hat{I}_{ef}^{\dagger} = \frac{1}{2\sqrt{2}}\bigg((-1)^{x_{j,e}}(\hat{c}_e^{\dagger}\hat{c}_f^{\dagger} - \hat{d}_e^{\dagger}\hat{d}_f^{\dagger} + \hat{c}_e^{\dagger}\hat{d}_f^{\dagger} - \hat{d}_e^{\dagger}\hat{c}_f^{\dagger}) -  (-1)^{x_{j,f}}(\hat{c}_f^{\dagger}\hat{c}_e^{\dagger} - \hat{d}_f^{\dagger}\hat{d}_e^{\dagger} + \hat{c}_f^{\dagger}\hat{d}_e^{\dagger} - \hat{d}_f^{\dagger}\hat{c}_e^{\dagger})\bigg)$ is the operator for the tuple $(e,f)$ that corresponds to incorrect parity outcome of the $e$-th and $f$-th bit of Bank's secret string $x_j$. {Comparing with the operator $\hat{T}^{\dagger}_{ef}$ in Eq~\ref{eq:C5}, we observe that the operator $\hat{I}_{ef}^{\dagger}$ has a negative sign  in the second term, which corresponds to an incorrect outcome. Operating $\hat{I}_{ef}^{\dagger}$ on the banknote state $\ket{x_j}$, results in 0 probability of an outcome since the banknote state always outputs a correct outcome. However, we use the fact that $p_{Ver1}^{ef,inc} \neq 0$ for any $\eta_{x_j}$ different from $\ket{x_j}\bra{x_j}$, and further this probability relates to how far these two states are in the metric of fidelity.}

Over all the tuples in $\mathcal{T}_n$, the probability that Ver1 obtains an incorrect outcome is,

\begin{equation}
\begin{split}
p_{Ver1}^{x_j, inc} &= \sum_{(e,f) \in \mathcal{T}_n}p_{Ver1}^{ef, inc} \\
&= \frac{1}{2n}\sum_{(e,f) \in \mathcal{T}_n}(A_{ee} + A_{ff} - (-1)^{x_{j,e} \oplus x_{j,f}}(A_{ef} + A_{fe})) \\
&= \frac{1}{2n}(n - \sum_{e,f}^{n}(-1)^{x_{j,e} \oplus x_{j,f}}A_{ef}) \\
&= \frac{1}{2}(1 - F_{x_j}) 
\end{split}
\label{D7}
\end{equation}

where $F_{x_j} = \bra{x_j}\eta_{x_j}\ket{x_j} = \frac{1}{n}\sum_{e,f}^{n}(-1)^{x_{j,e} \oplus x_{j,f}}A_{ef}$ is the square of fidelity between the adversary's state $\eta_{x_j}$ and the banknote state $\ket{x_j}$. For simplicity we refer to this squared fidelity as fidelity.

The above probability is calculated for a specific string $x_j$. Since the adversary does not know this string, she instead holds a state which is a mixture of all possible $2^n$ strings $x_j$. Thus the averaged out error probability for Ver1 is,

\begin{equation}
\begin{split}
p_{Ver1}^{inc} &= \frac{1}{2^n}\sum_{x_j} p_{Ver1}^{x_j} = \frac{1}{2}(1 - F)
\end{split} 
\end{equation}

where $F = \frac{1}{2^n}\sum_{x_j}F_{x_j}$. 

Similar analysis for Ver2, who receives the mixed state $\tau_{x_j}$, the fidelity with the honest note state is $G_{x_j} = \bra{x_j}\tau_{x_j}\ket{x_j}$. The average error probability of obtaining an incorrect outcome is $p_{Ver2}^{inc} = \frac{1}{2}(1 - G)$, where $G = \frac{1}{2^n}\sum_{x_j}G_{x_j}$. We cast the objective problem for the adversary to minimize the  average error probability of Ver1 and Ver2,

\begin{equation}
p_{Ver1}^{inc} + p_{Ver2}^{inc} = 1 - \frac{F+G}{2}
\label{eq:comberr}
\end{equation}

This minimization problem can be cast as a semi-definite program (SDP) with the objective to find a lower bound of Eq~\ref{eq:comberr}. This can alternatively be viewed as maximizing the average fidelity $\bar{F} = \frac{F + G}{2}$. 

{The security proof in quantum money proposal by Amiri et al. \cite{amiri2017quantum} has a similar reduction of the average error probability to maximizing the average fidelity of the states received by the two verifiers, Ver1 and Ver2, with the original banknote state. Thus one is interested in finding the completely-positive trace-preserving physical channel that takes a valid banknote state and prepares one copies for Ver1 and Ver2 with maximal average fidelity. Translating this problem in an SDP formalism requires using the Choi-Jamiolkowski representation \cite{jamiolkowski1972linear} that translates the channel into a corresponding state in a higher dimension using maximally entangled state $\ket{\psi} = \sum_{i=1}^{n}\ket{1}_{i}\ket{1}_i$. Here we directly use the SDP result of \cite{amiri2017quantum}. Further details on this SDP construction can be found directly in their work. They numerically verified that for $n \leqslant 14$, }   

\begin{equation}
\bar{F} \leqslant \frac{1}{2} + \frac{1}{n}
\label{Eq:D9}
\end{equation}  

{This implies that the average fidelity of the states received by Ver1 and Ver2 with the banknote state is upper bounded by a factor less than 1 for $n \geqslant 4$. Further, the upper bound on average fidelity decreases with increasing $n$ and if Eq~\ref{Eq:D9} is true for any $n$ (conjectured by \cite{amiri2017quantum}), then for $n \gg 4$, any optimal strategy by the adversary results in average fidelity that is upper bounded by 0.5. } Eq~\ref{Eq:D9} allows us to give a lower bound on the average probability that Ver1 and Ver2 obtains an incorrect outcome of a tuple in $\mathcal{T}_n$, 

\begin{equation}
\begin{split}
p_{Ver1}^{inc} + p_{Ver2}^{inc} &= 1 - \frac{1}{2}(F + G) \\
&\geqslant \frac{1}{2} - \frac{1}{n} 
\label{Eq:D10}
\end{split}
\end{equation}

This is the probability for a single copy $j$ chosen by the verifier. Across each of the chosen $L$ uniformly random copies, the average error probability is lower bounded by Eq.~\ref{Eq:D10} in the collective adversary attack scenario.  
Since the above equation gives us a lower bound on the average error probability for both verifiers, this implies that the minimum error probability $e_{min}$ for any one of the two verifiers, lets say Ver1, must definitely be,

\begin{equation}
e_{min} = \frac{1}{4} - \frac{1}{2n}
\end{equation}

The above error probability has been calculated for $q' = q - (3T-4)|L|$ copies. Over the remaining $q - q' = (3T-4)|L|$ copies, we assume that the adversary has full information of the state. Hence the error probability in obtaining the parity outcome of a tuple for the $q - q'$ copies is 0 for Ver1 (same for Ver2). The minimum error probability for Ver1 averaged across the $l_{succ}$ copies (where single clicks have been obtained across two distinct modes) is,

\begin{equation}
e_{min} = \frac{q-(3T-4)|L|}{q-(T-1)|L|}\big(\frac{1}{4} - \frac{1}{2n}\big) 
\label{Eq:VerForge}
\end{equation}

where the denominator $q-(T-1)|L|$ is due the fact that the adversary can set $(T-1)|L|$ positions in the $r$ register to 1, thus effectively ensuring that Ver1 does not select the note copies in those positions. Suppose $T|L| = \lambda q$, for some small fraction $\lambda < 1$ (for example $1/1000$), then Eq.(\ref{Eq:VerForge}) is,

\begin{equation}
e_{min} \approx \frac{997}{999}\big(\frac{1}{4} - \frac{1}{2n}\big) \approx \frac{1}{4} - \frac{1}{2n}
\label{Eq:VerForge1}
\end{equation}

We know that if the holder is honest, the probability of him obtaining the correct parity outcomes across all the $l_{succ}$ copies is $c=1$. From Eq.~\ref{Eq:VerForge1}, we see that the corresponding probability of obtaining correct parity outcomes for the adversary is upper bounded by $c_{\text{adv}} = 1 - e_{min}$. {The gap $c - c_{\text{adv}}$ is defined as the noise tolerance of our protocol since it denotes the maximum theoretical probability that an honest noteholder returns an incorrect outcome while still maintaining a non-negative gap in the success probability with any strategy by the adversary. Denoting the cut-off $\delta = (c - c_{\text{adv}})/2$ and using the Chernoff-Hoeffding bound \cite{yao1983lower}, we can now compute the probability that adversary's note passes the test of \emph{Verification Phase} by both Ver1 and Ver2 is, }

\begin{equation}
\begin{split}
\mathbb{P}[\text{Adv pass}]&= \mathbb{P}[\text{Ver1 and Ver2 accept}]\times \mathbb{P}[\text{Ver1}^{B}_{H} = 1 \hspace{1mm}\text{and}\hspace{1mm} \text{Ver2}^{B}_{H} = 1 |\text{Ver1 and Ver2 accept}] \\
&\leqslant \mathbb{P}[\text{Ver1 accepts}]\cdot\mathbb{P}[\text{Ver1}^{B}_{H} = 1|\text{Ver1 accepts}] \\
&\leqslant \mathbb{P}[\text{Ver1}^{B}_{H} = 1|\text{Ver1 accepts}]\\
&\leqslant \exp(-\frac{2\delta^2 l_{min}^2}{|L|}) = \exp(-2\delta^2 p_{11}^2 (1 - \epsilon)^2 |L|)
\end{split}
\end{equation}

The condition $c > c_{\text{adv}}$ always holds as long as $n > 2$, hence the probability that both verifiers pass the verification test is exponentially low. Since the Eq.~\ref{Eq:D9} has been verified until $n= 14$, the maximum noise tolerance of the scheme is up to $21.4\%$. Further, if Eq.~\ref{Eq:D9} is true for all $n$, the the maximum noise asymptotic noise tolerance of $25\%$ can be achieved with our scheme.

{\textbf{3.2 Coherent attack:} The collective attack strategy focuses on the optimal manipulation of individual copies of the note state. However, the adversary can perform a combined operation on all the copies of the input of state to potentially decrease the average error probability on each copy of the note state lower than the bound Eq.~\ref{Eq:D10}. Alternatively, the adversary can hope for a general entangled strategy on the $T$ chosen copies of the verifiers such that conditional on the measurement outcomes obtained in the first $T-1$ copies, the value of average error probability for the last copy is decreased even below the bound of Eq.~\ref{Eq:D10}. To mitigate any such strategy, we use the results of Croke and Kent \cite{croke2012security} which proves the security of a quantum relativistic bit commitment protocol against any adversary. Their reduction of a coherent attack strategy to a collective strategy uses the teleportation based argument to show that any error probability achieved via coherent strategy can also be achieved by individual manipulation of the states by an adversary who uses a maximally mixed entangled state to teleport the original state to the verifier (along with the teleportation correction). Also since the collective attack is optimal manipulation of individual states, hence it also includes teleportation based strategies. Our setting is very similar to their construction and allows us to use their results in a straightforward manner. Therefore, any coherent strategy cannot beat the lower bounds proved in the collective strategy. } \par

{\textbf{4. Measure and Resend attack:} Till now we have analysed a generic way to upper bound the success probability in any adversary attack scenario. To gain intuition on how an adversary attack would play out, we consider a specific example attack called `measure and resend'. Here for simplicity, we consider that the adversary does not perform register manipulation or adaptive attack, but rather she simply measures each copy of the incoming state and creates two states based on the measurement outcome to be sent to Ver1 and Ver2. }

{Assume that the Bank sends a note with $q=10^6$ copies of single-photon states to the holder. Each copy is encoded with Bank's secret string $x_j \in \{0,1\}^4$ for $j \in [10^6]$. For verification, $|L| = 10^3$ copies of the note states are chosen at random and Sampling Matching scheme is performed on each of the copies to obtain a parity measurement outcome.}

{Suppose the strategy by an adversary is to measure each banknote copy $\ket{x_j}$ in the basis},
\begin{equation}
\bigg\{\frac{1}{\sqrt{2}}(\ket{1}_1 \pm \ket{1}_2), \hspace{2mm} \frac{1}{\sqrt{2}}(\ket{1}_3 \pm \ket{1}_4)\bigg\}
\end{equation}
{With this strategy, Bob can always provide the correct parity outcome either the tuple $\{1,2\}$ or $\{3,4\}$. This is because the outcome $\frac{1}{\sqrt{2}}(\ket{1}_1 + \ket{1}_2)$ only occurs if $x_{j,1} \oplus x_{j,2} = 0$, while $\frac{1}{\sqrt{2}}(\ket{1}_1 - \ket{1}_2)$ only occurs if $x_{j,1} \oplus x_{j,2} = 1$. Similarly the outcomes $\frac{1}{\sqrt{2}}(\ket{1}_3 \pm \ket{1}_4)$ provide a prarity outcome for the tuple $\{3,4\}$. Note that since the state $\ket{x_j}$ contains $\log_2 n$ qubits/bits of information, any strategy by the adversary would limit her information retrieval to only $\log_2 n$ bits. This is a consequence of the Holevo's bound \cite{holevo1973information} which states that it is impossible to retrieve more than $\log_2 n$ bits of information from a state of the form $\ket{x_j}$. Thus the adversary cannot retrieve the entire information of the secret string $x_j$ and hence she cannot forge the note perfectly.}

{Using the above strategy, the adversary gets the parity information of exactly one tuple. In the \textbf{resend} phase to Ver1 and Ver2, he  correctly encodes that parity information, while for the rest bits of which he has no information, he randomly encodes them in 0 or 1. As an example, for the $j$-th copy, if the adversary get the information of the tuple $\{1,2\}$, the state she sends to both verifiers Ver1 and Ver2 is
\begin{equation}
    \ket{\text{adv}} = \frac{1}{\sqrt{2}}((-1)^{r_1}\ket{1}_1 + (-1)^{r_2}\ket{1}_2 + (-1)^{r_3}\ket{1}_3 + (-1)^{r_4}\ket{1}_4)
\end{equation}
where he picks $r_1, r_2$ such that $r_1 \oplus r_2 = x_{j,1}\oplus x_{j,2}$, and $r_3, r_4$ are picked at random from 0/1. Upon receiving the note, Ver1 (similarly Ver2) runs the Sampling Matching scheme with his local $\ket{\beta} = \frac{1}{2}\sum_{i=1}^{4}\ket{1}_i$ on the randomly chosen $10^3$ copies of the note state. From Eq.\ref{2photonsamemode} and \ref{2photonsamemodeAdv}, we see that across $p_2\times 10^3 = 250$ copies, Ver1 will receive 2 photon clicks in the same mode on average thus not obtaining the parity information of any tuple. Across the remaining $750$ copies on average, he receives the parity outcome for any one of the three tuples \{$\{1,2\}, \{1,3\}, \{2,3\}\}$ with equal probability. Since the adversary managed to successfully extract only a single parity outcome, so the probability that the parity outcome obtained by Ver1 is wrong would be,
\begin{equation}
    p_{Ver1}^{x_j,inc} = \sum_{\text{tuple}}p_{\text{tuple}} \times  p_{\text{correct}|\text{tuple}} = \frac{1}{3}(0 + \frac{1}{2} + \frac{1}{2}) = \frac{1}{3}
\end{equation}
since the adversary will be correct in one parity outcome and the random guessing for remaining two parity outcomes gives him a success rate of 0.5. Thus the sucess probability of the adversary on each copy is $c_{\text{adv}} = 2/3$. If the noteholder was honest, this success probability $c = 1$ since the banknote was not tampered with. Denoting the cut-off $\delta = (c - c_{\text{adv}})/2 = 1/6$ and using Chernoff-Hoeffding bound, the probability that adversary's note passes the test of Verification phase (whene Ver1 communicates the parity outcome to the Bank) is,
\begin{equation}
\begin{split}
     \mathbb{P}[\text{Adv pass}]&=  \mathbb{P}[\text{Ver1 accepts}]\cdot\mathbb{P}[\text{Ver1}^{B}_{H} = 1|\text{Ver1 accepts}] \\
&\leqslant \mathbb{P}[\text{Ver1}^{B}_{H} = 1|\text{Ver1 accepts}]\\
&\leqslant \exp(-2\delta^2 (1 - p_{2}) |L|)  \\
&\leqslant \exp(-41) \approx 10^{-17}
\end{split}
\end{equation}
Thus we see that it is virtually impossible for an adversary to forge the note with this message and resend attack. }

\subsection{Quantum Money Scheme with Coherent states} \label{coherentstates}

In this section, we briefly describe the private quantum money scheme when the Bank encodes secret strings into weak coherent states instead of the single-photon states. The primary reason we want to encode the note as coherent states is that it facilitates the implementation of Sampling Matching in a much simpler and elegant manner. With coherent states, as shown by Kumar et al. \cite{kumar2019experimental}, the Sampling Matching implementation requires a single 50/50 beam splitter and two single-photon threshold detectors irrespective of input size of the strings. 

In the coherent state encoding, the Bank independently and randomly chooses $q$ $n$-bit binary strings $x_1,x_2,..,x_q \in \{0,1\}^n$. Each string $x_j$ is now encoded into the coherent state $\ket{\alpha_{x_j}}$, with an average photon number 1,

\begin{equation}
\ket{\alpha_{x_j}} = \bigotimes_{k=1}^{n} \ket{(-1)^{x_{j,k}}\frac{1}{\sqrt{n}}}_k
\end{equation}

where $x_{j,k}$ is the $k^{th}$ bit value of string $x_j$. The coherent state $\ket{\alpha_{x_j}}$ is a sequence of $n$ coherent pulses in $n$ modes. The banknote is then the sequence of $q$ coherent states,

\begin{equation}
\$ = \bigotimes_{j=1}^{q} \ket{\alpha_{x_j}}
\end{equation}  

This is then distributed among the untrusted noteholders. To carry out a transaction, the holders send the note to the local verifiers. The verifier uniformly and randomly selects few copies of the note state to run the Sampling Matching scheme. In this scheme, for each selected copy $\ket{\alpha_{x_j}}$, the verifier prepares his local state $\ket{\beta} = \bigotimes_{k=1}^{n} \ket{\frac{1}{\sqrt{n}}}_k$ as a sequence of $n$ coherent pulses. This state is sequentially interacted with the selected states of the note chosen by the verifier. The interaction is via the 50/50 beam splitter. The coherent pulse modes at the input of verifier's beam splitter in $k$-th step are,

\begin{equation}
\ket{(-1)^{x_{j,k}}\frac{1}{\sqrt{n}}}_k \otimes \ket{\frac{1}{\sqrt{n}}}_k,
\label{eq:in}
\end{equation}

and the output modes are,

\begin{equation}
\ket{\frac{(1+(-1)^{x_{j,k}})}{\sqrt{2}}\frac{1}{\sqrt{n}}}_{k,D_0} \otimes \ket{\frac{(1-(-1)^{x_{j,k}})}{\sqrt{2}}\frac{1}{\sqrt{n}}}_{k,D_1}
\label{eq:out}
\end{equation}

The output modes are fed into the single-photon threshold detectors $D_0$ and $D_1$ to observe the clicks. When a coherent state $\ket{\alpha}$ is incident on the threshold detector, the probability of the click is given by,

\begin{equation}
p_c = 1 - \exp(-|\alpha|^2)
\end{equation}

\begin{figure}[t!]
\includegraphics[scale=0.55]{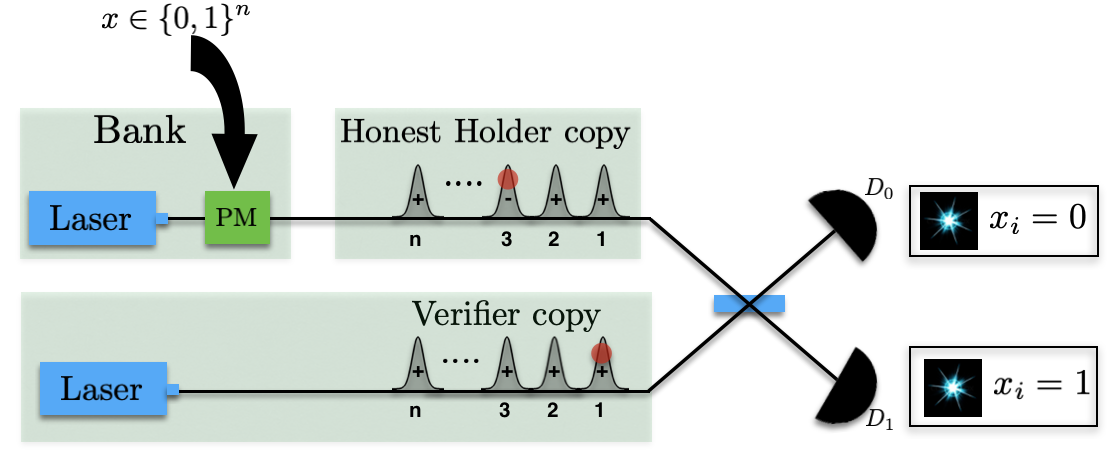}
\centering
\caption{Sampling Matching (SM) circuit implementation using weak coherent states, 50/50 beam splitter (BS) and single-photon threshold detectors. The upper arm illustrates an honest holder's state as a coherent state $\ket{\alpha_x}$, which consists of a sequence of coherent pulses. The coherent state is encoded with a random phase $x \in \{0,1\}^n$ through the phase modulator (PM). The lower arm is used by the verifier to produce a local reference coherent state $\ket{\beta}$, consisting of a sequence of coherent pulses, with an average photon number of 1. The verifier interferes the states into the 50/50 BS and infers the parity information from the detector clicks in $D_0$ and $D_1$. This allows him to obtain the parity outcome of a tuple in $\mathcal{T}_n$. The red dot in the $1^{st}$ and $3^{rd}$ time sequence denotes that the verifier observed clicks at $D_1$ and $D_0$ detectors respectively for these time steps. Thus he infers the parity outcome for the tuple $(1,3)$,  $x_1 \oplus x_3 = 1$.} 
\label{fig:qmsm3}
\end{figure}
Figure~\ref{fig:qmsm3} is a depiction of the sequential interaction of the coherent pulses of one copy of an honest noteholder and the verifier's local state. 

Let us see how the verifier obtains the parity outcomes of one of the tuples in $\mathcal{T}_n$ from the detector clicks.  The output state in Eq.(\ref{eq:out}) denotes that the detector $D_0$ clicks iff $x_{k}  = 0$  while $D_1$ clicks iff $x_{k} = 1$. For each of the chosen copy $j$, the verifier will be unable to infer the parity outcome of any matching with certainty if he does not obtain single-click in atleast two time steps. This probability $\mathbb{P}$(no two single-clicks) = $\mathbb{P}$(no single-clicks) + $\mathbb{P}$(exactly one single-click). We denote this probability by $p_{\neg 11}$,

\begin{equation}
p_{\neg 11} = (1 - p_1)^n + {n\choose 1}p_1(1 - p_1)^{n-1}
\end{equation}

where $p_1 = 1 - \exp(-\frac{2}{n})$ is the probability of observing a single click in one time step. Thus with probability $1 - p_{\neg 11}$, which grows with $n$, the verifier obtains two single-photon clicks in the chosen state. 

Now suppose the verifier observes single-photon clicks in the $k$-th and $l$-th time modes in detectors $D_0$ and $D_1$ respectively. This implies $d = x_{k} \oplus x_{l} = 1$. This enables the verifier to output the parity outcome of $(k,l) \in \mathcal{T}_n$. If on the other hand, the verifier does not obtain exactly two clicks in two different time modes, he outputs the parity outcome $d = \varnothing$.

\subsubsection{Description of the Money Scheme}

\begin{figure}[h!]
\includegraphics[scale=0.52]{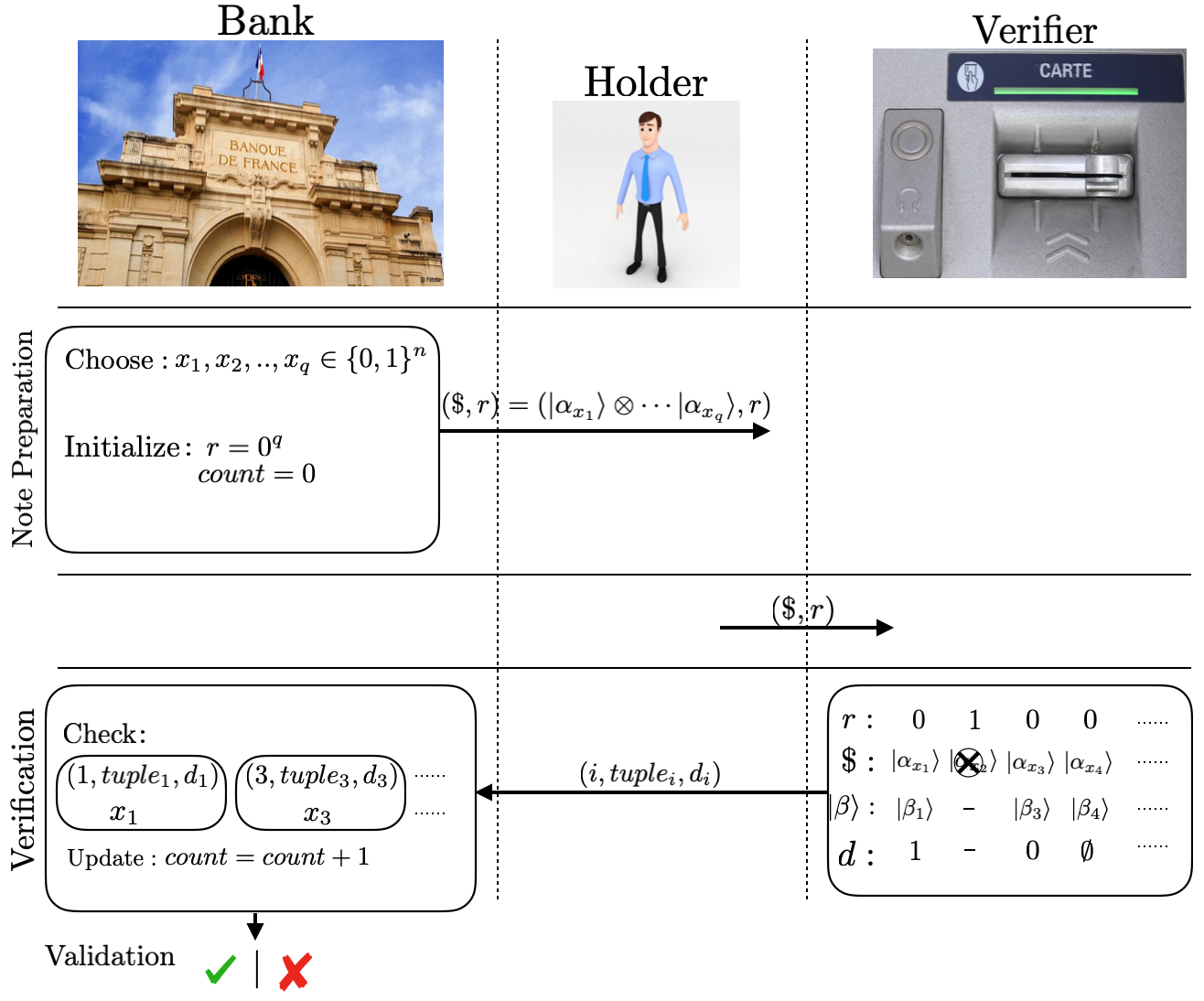}
\centering
\caption{Illustration of our quantum money scheme based on the verification protocol using the SM-scheme. In the \emph{
Note Preparation} phase, the Bank independently and randomly selects $q$ n-bit binary strings and produces note coherent states $\$ = \ket{\alpha_{x_1}}\otimes\ket{\alpha_{x_2}}\otimes\cdots\otimes\ket{\alpha_{x_q}}$. The Bank initializes the $r$ register to keep a track of the number of positions in $[q]$ where the states have been used for verification and the $count$ register to keep track of the number of verification attempts on the note. The note $(\$,r)$ is then sent to the holder. To carry out a transaction, the holder sends the note to an honest verifier of the Bank. In the \emph{Verification} phase, the verifier selects a fraction of the $q$ copies of the note state with positions 0 in the $r$ register. He creates his local state $\ket{\beta_j}$ and applies the SM-scheme on those selected copies. If the note passes the local test of the verifier, he sends the measurement outcomes of the test to the Bank. Finally, the Bank compares the outcomes with his secret string $x_j$'s  and outputs a bit $\text{Ver}^{B}_{H}$ stating whether the note is valid or not. }
\label{fig:SHMPic}
\end{figure}

We divide our quantum money scheme using coherent states into two phases. First is the \emph{Note preparation phase}, where the Bank chooses multiple n-bit binary strings independently and randomly. The Bank takes each of these individual strings to  produce weak coherent states. The quantum note $\$$ of the Bank is the combined tensor product of coherent states corresponding to all the input strings. This is distributed among the untrusted holders. In the \emph{Verification phase}, a noteholder sends the note to the verifier in order to carry out a transaction. Upon receiving the note, the verifier randomly selects some copies of the note state (here the note consists of multiple copies, where one copy corresponds to the coherent state that encodes one n-bit string). For the selected copies, he runs the verification protocol using the Sampling Matching, SM scheme (Figure~\ref{fig:qmsm3}). He locally checks if the statistics of the measurement outcome obtained by running the SM-scheme is what he should expect from an honest noteholder. If he finds discrepancies, the note is rejected. If the note passes this test, then the outcomes from the SM-scheme are classically communicated with the Bank. The Bank compares these outcomes with his private $n$-bit strings. If a high fraction of the outcomes is correct, he outputs the bit $\text{Ver}^{B}_{H} =1$ implying that the note is valid. Otherwise, he outputs $\text{Ver}^{B}_{H} = 0$.  

\subsubsection{Note Preparation Phase}

\begin{enumerate}
  \item The Bank independently and randomly chooses $q$ $n$-bit binary strings $x_1,x_2,..,x_q \in \{0,1\}^n$ 
  \item The Bank encodes each the binary string $x_j$ into the phase randomized coherent state $\ket{\alpha_{x_j}}$, with an average photon number 1,
  
  \begin{equation}
  \ket{\alpha_{x_j}} = \bigotimes_{k=1}^{n} \ket{(-1)^{x_{j,k}}\frac{1}{\sqrt{n}}}_k
  \end{equation}
  
  where $x_{j,k}$ is the $k^{th}$ bit value of string $x_j$. The coherent state $\ket{\alpha_{x_j}}$ is a sequence of $n$ coherent pulses in $n$ modes. {The note is a combination of all the $q$ copies of the coherent state $\$ = \ket{\alpha_{x_1}}\otimes\ket{\alpha_{x_2}}\otimes\cdots\otimes\ket{\alpha_{x_q}}$.}
 \item The Bank creates a classical binary register $r$ and initializes it to $0^q$. This register keeps the track of positions $j$ where the states  have been used for the verification.
 \item The Bank  creates a counter variable $count$ and initializes it to 0. This keeps a track of the number of verification attempts. 
 \item  The Bank sends the quantum note ($\$$, r) to the holder.\\
\end{enumerate} 

\subsubsection{Verification Phase}

Once the Bank distributes the notes, the holder in order to be able to carry out any transaction, has to get the note verified from an honest verifier Ver. The verification procedure is listed below. 
\\\\
\textbf{Local testing}
\begin{enumerate}
\item The holder gives the note $\$' (=: \$$ if the holder is honest) to Ver. 
\item Ver checks the re-usability of the note by verifying that the hamming distance of $r$ register $d(r,0^q) \leqslant T$, where $T$ is a predefined maximum number of copies in the note that are allowed for verification. If $d(r,0^q) > T$, the note is rendered useless and must be returned to the Bank.
\item Ver uniformly and randomly selects a subset $L \subset [q]$ copies from the states marked $r = 0$. He marks all the corresponding $|L|$ copies in the $r$ register to 1.
\item For each copy $j \in L$, Ver prepares his local coherent state $\ket{\beta_j} = \bigotimes_{k=1}^{n}\ket{\frac{1}{\sqrt{n}}}_k$ and runs the SM scheme (Figure~\ref{fig:qmsm3}).
\item Ver counts the number of successful copies $l_{succ}$, where he obtains exactly two single-photon clicks in two different time modes. For these copies he outputs the parity outcome  $d_j = x_{j,k} \oplus x_{j,l}$ where the clicks have been obtained in times modes $k$ and $l$. For the rest of the copies, he sets $d_j = \varnothing$.
\item Ver checks if $l_{succ} \geqslant l_{min}$, where $l_{min} = \mathbb{E}_{h}[l_{succ}](1 -\epsilon)$ is the  minimum number of copies that will locally guarantee his acceptance of the note, where $0\leqslant\epsilon\leqslant 1$ is the security factor. Here $\mathbb{E}_{h}[l_{succ}]$ is the expected number of copies where the honest noteholder obtains exactly two single-photon clicks when Ver runs the SM scheme.
\item Ver proceeds to the communication with the Bank only when the note passes this test.
\end{enumerate}
\textbf{Communication with the Bank}
\begin{enumerate}
\item[8.] Ver forwards the outcomes $\{j \in L,(k,l),d_j\}$ to the Bank.
\item[9.] The Bank checks if $count < \lceil \frac{T}{|L|} \rceil$, otherwise he renders the verification attempt as invalid.
\item[10.] For each copy $j \in L$ with $d_j \neq \varnothing$, the Bank compares the parity value $d_j$ with the secret string $x_j$. He validates the note if the number of correct outcomes

\begin{equation}
l_{succ}^{cor} \geqslant \mathbb{E}_h[l_{succ}^{cor}](1 - \delta)  
\end{equation}

where $\mathbb{E}_h[l_{succ}^{cor}]$ is the expected number of copies that give the correct parity outcome when the noteholder is honest, and $0\leqslant\delta\leqslant 1$ is a positive constant.
\item[11.] The Bank updates the $count$ by 1.
\end{enumerate}

Here we are not providing full security proof of our quantum money scheme using coherent states. This is due to the fact that the statements of security would be dependent on the specific experimental parametric values. The main objective of this section is to illustrate the simplicity and elegance of mapping the money scheme using coherent states which simplifies the verification circuit requirement to only a $\mathcal{O}(1)$ 50/50 beam splitters and two single-photon threshold detectors. Nevertheless, we expect that a full security proof can be constructed in a straightforward manner from the security proof given for the single-photon superposition states.


\section{Discussion}

We have introduced the private quantum money scheme as a cryptographic task using the Sampling Matching verification scheme. Sampling Matching is an experimentally motivated verification framework to ease out the implementation of quantum money schemes. Our proposed money scheme demonstrates an information-theoretic security against any adversary with a noise tolerance of $21.4\%$. We mostly focus on the scheme with the banknotes being single-photon states. This has been done keeping in mind that single-photon states are a direct realisation of qubit/qudit with linear optics, thus our money scheme definition can be directly translatable to other qubit encoding pictures. The security analysis of our money scheme considers the most general attack by the adversary trying to duplicate the banknotes which can be passed by the Bank with high probability. In the second half, we have proposed a money scheme to encode the secret strings into a sequence of coherent states with the verification using Sampling Matching consisting of a single 50/50 beam splitter and 2 single-photon threshold detectors. This scheme provides a dramatic reduction in the component requirements by the verifier thus experimentally facilitate achieving higher noise tolerance than what is realistically feasible with other quantum money schemes. 

Another major challenge in quantum money schemes is the storage of quantum states of the Bank in a quantum memory which can then be distributed to the holder. There has been considerable ongoing works in the development of quantum memory for storing and retrieving the quantum data \cite{lvovsky2009optical, julsgaard2004experimental, fleischhauer2002quantum, kozhekin2000quantum}. Even though this is not the specific focus of our work as our approach is to simplify the verifier's circuit, we still address the above memory issue partially by introducing the verifier circuit (when the banknotes are encoded coherent states) that works even when the states are sent on the fly i.e. no memory is required by the verifier to store them. This is due to the fact that the coherent state encoding is a tensor product of individual coherent pulses and the beam splitter interaction with the verifier's state occurs separately on each of these pulses.

Finally, the verification tool using Sampling Matching is a generic tool that has the potential to be applied to other cryptographic tasks such as key-distribution, and also in quantum verification protocols such as efficient verification of prover's state with limited verifier resources in scenarios where the prover does not want to fully reveal the secret encoding \cite{arrazola2018quantum}. Since most cryptographic schemes rely on one-way functions and the fact that it is hard for an adversary to invert the function without having the knowledge of the secret key that was used to prepare the function, the scheme using Sampling Matching offers an excellent candidate to be such an experimentally realisable function.

\medskip
\bibliographystyle{alpha}
\bibliography{fingerprints}

\end{document}